%% file: ANA-STDM-2017-05-PAPER.tex
\documentclass[PAPER,texlive=2023, cernpreprint, UKenglish, texmf, orcidlogo]{atlasdoc}
 
\usepackage{atlaspackage}
\usepackage{atlasbiblatex}
 
\usepackage{atlasjournal}
\usepackage{atlasmisc}
\usepackage{atlasparticle}
\usepackage{atlasunit}
\usepackage{atlasxref}
\usepackage{atlasphysics}

\hypersetup{colorlinks,breaklinks}
\hypersetup{linkcolor=blue,citecolor=blue,filecolor=black,urlcolor=blue}
\hypersetup{pdftitle={ATLAS draft cover},pdfauthor={The ATLAS Collaboration}}
 
\usepackage{atlascontribute}

\graphicspath{{logos/}{figures/}}
 
\usepackage{ANA-STDM-2017-05-PAPER-defs}
\newcommand{\Zllg}{\ensuremath{Z \to ll\gamma}}
\newcommand{\Zmmg}{\ensuremath{Z \to \mu\mu\gamma}}
\newcommand{\Zeeg}{\ensuremath{Z \to ee\gamma}}
\newcommand{\Zllgg}{\ensuremath{Z \to ll\gamma\gamma}}

\AtlasTitle{Study of \Zllg\ decays at $\sqrt s$~= 8 TeV with the ATLAS
detector}

\AtlasVersion{2.0}
 
\AtlasAbstract{
This paper presents a study of \Zllg~decays with the ATLAS detector at
the Large Hadron Collider. The analysis uses a proton--proton data sample
corresponding to an integrated luminosity of 20.2~fb$^{-1}$
collected at a centre-of-mass energy $\sqrt{s}$ = 8 TeV. Integrated
fiducial cross-sections together with normalised differential fiducial
cross-sections, sensitive to the kinematics of final-state
QED radiation, are obtained.  The results are found to be in
agreement with state-of-the-art predictions for final-state QED
radiation. First measurements of \Zllgg~decays are also reported.
}

\author{The ATLAS Collaboration}
 
\AtlasRefCode{STDM-2017-05}

\PreprintIdNumber{CERN-EP-2022-265}
 
\AtlasJournal{EPJC}
\AtlasJournalRef{Eur. Phys. J. C 84, 195 (2024)}
\AtlasDOI{10.1140/epjc/s10052-024-12471-9}

\hypersetup{pdftitle={ATLAS document},pdfauthor={The ATLAS Collaboration}}

\begin{document}
 
\maketitle
 
\tableofcontents
 
\section{Introduction}
\label{sec:Intro}

The production of $Z$~bosons and their decay to lepton pairs at the Large Hadron Collider~(LHC) through the Drell-Yan mechanism has been a topic of very fruitful and detailed studies in the LHC~experiments, and the precision of the measurements has led over the past decade to impressive theoretical developments, mostly in the area of higher-accuracy quantum chromodynamic~(QCD)~predictions.
 
\begin{figure}[htb!]
\begin{center}
\includegraphics[width=0.47\textwidth]{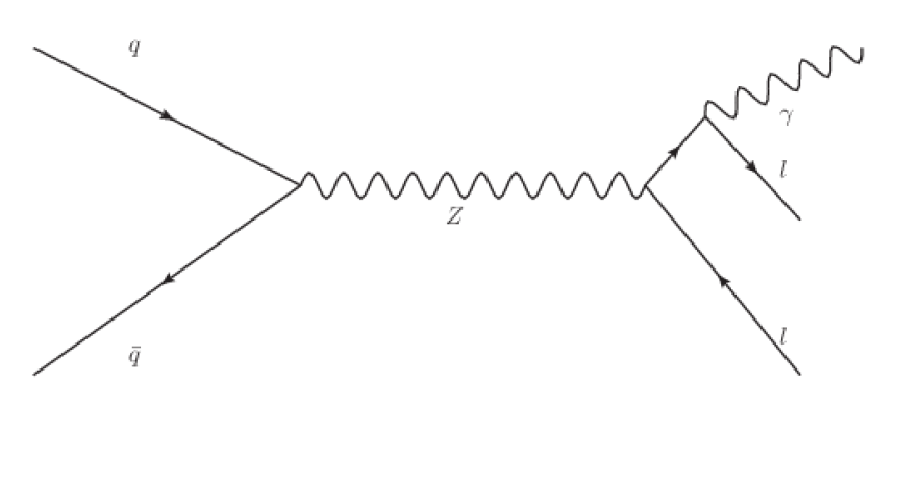}
\hfill
\includegraphics[width=0.47\textwidth]{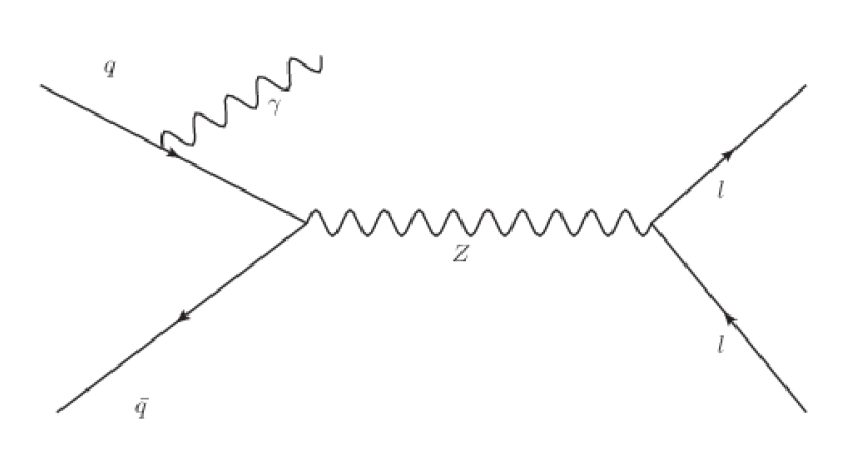}\\
\parbox[t]{0.47\textwidth}{\caption{Example of $q\bar{q} \to Z \to ll\gamma$
final-state QED radiation diagram.\label{fsr}}}
\hfill
\parbox[t]{0.47\textwidth}{\caption{Example of $q\bar{q} \to Z \to ll\gamma$
initial-state QED radiation diagram.
\label{isr}}}
\end{center}
\end{figure}

Precise predictions for quantum electrodynamic and
electroweak~(QED/EW) effects are also of prime importance for key
measurements at the~LHC, such as that of the $W$-boson mass or of the
weak mixing angle; this became very clear already at the end of the
Run-1 data-taking period~\cite{WmassATLAS}. The analysis described in
this paper focuses on the final-state radiation of photons in
Drell-Yan production of $Z$~bosons decaying to an electron or muon
pair. This process, denoted in the following as QED~FSR~(Fig.~\ref{fsr}), can be
calculated separately from QED initial-state radiation or QED ISR~(Fig.~\ref{isr}) and from
initial-final-state interference (IFI) thanks to gauge invariance.
Owing both to the kinematic selections applied to the final state
particles, and to the narrow width of the $Z$~boson, the contributions
from QED~ISR and QED~IFI in the fiducial phase space of the
measurements described below are expected to be very
small~\cite{KKMChh20IFI}.
 
\begin{figure}[htb!]
\begin{center}
\includegraphics[width=0.48\columnwidth]{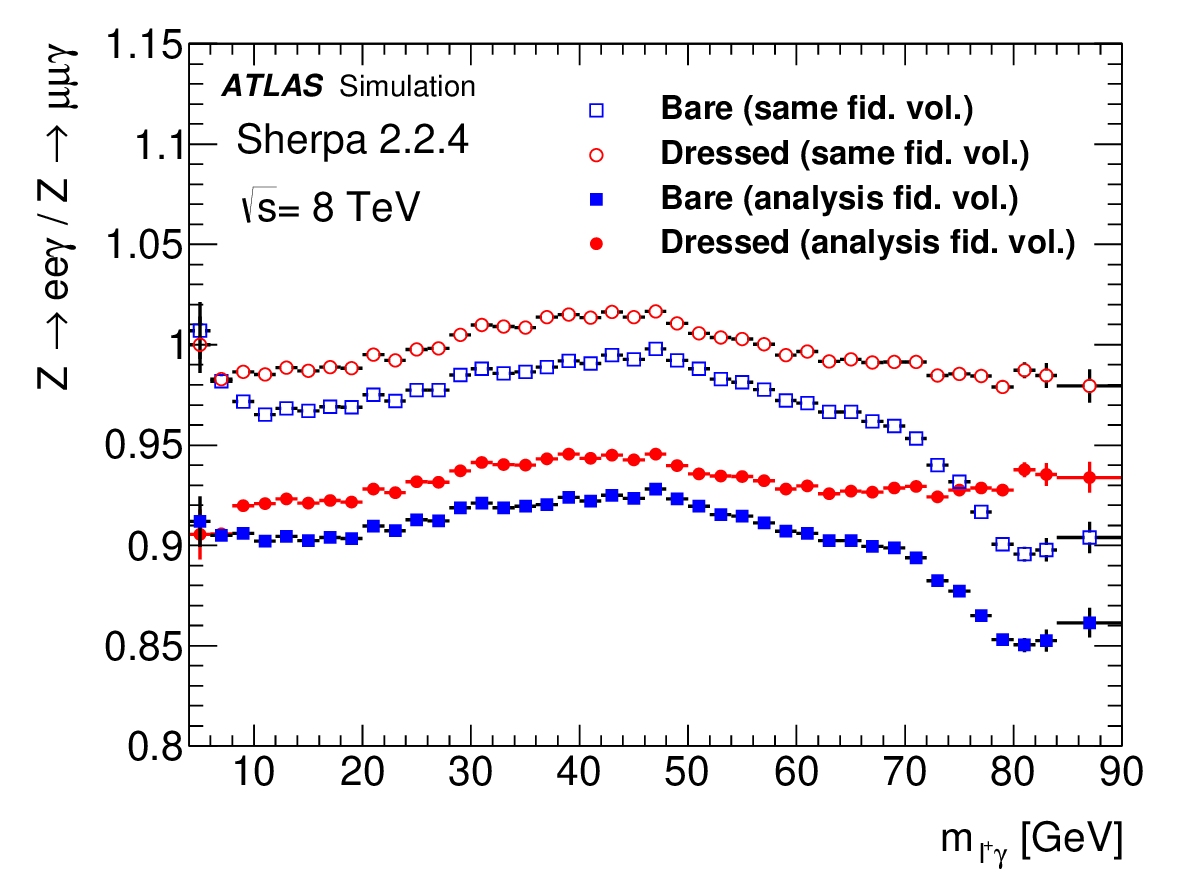}
\hfill
\includegraphics[width=0.48\columnwidth]{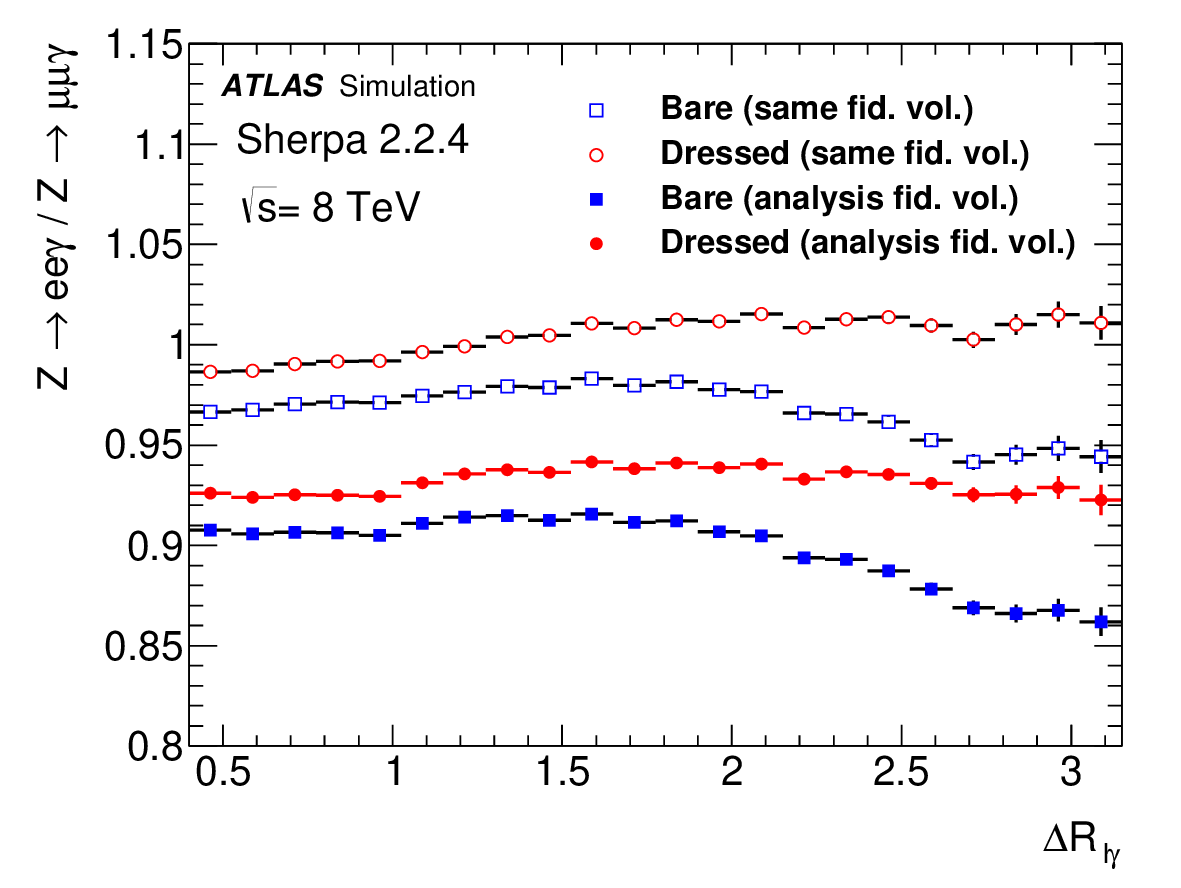}\\
\caption{
Predicted ratios of expected differential cross-sections for the invariant mass,~\mlg, of the positively charged lepton and the photon (left) and of the angular separation,~\DeltaRlg, between the photon and the closer of the two leptons (right) for electrons compared to muons. The ratios are shown for the bare and dressed lepton definitions described in the text. The predictions are obtained from a high-statistics inclusive $Z$-boson sample using Sherpa~2.2.4 and are shown for events containing two high-\pt leptons and one high-\pt photon satisfying the fiducial volume selections described in~Table~\ref{cutsTrX} in~Section~\ref{sec:unf} (full symbols). Since most of the difference between electrons and muons observed in this case for the bulk of the distributions is due to the higher acceptance in~$\eta$ for the muons, the electron to muon ratios are also shown in the case of identical fiducial selections for both charged leptons (open symbols).
\label{Sherpa_evsmu}
}
\end{center}
\end{figure}
 
When considering QED radiation, fixed-order partitioning of the
calculations has been shown at the time of the Large Electron-Positron
Collider~(LEP) to be sub-optimal in terms of obtaining the most
precise predictions.
 
Photon resummation techniques using exponentiation were originally
developed by Yennie-Frautschi-Suura~(YFS)~\cite{YFS} for inclusive
QED~FSR in the case of $e^+e^-$~collisions. They were then expanded to
be used for exclusive QED~ISR and~FSR and for $pp$~collisions at
the~LHC~\cite{KKMChh20IFI}. This analysis uses and compares several
state-of-the-art implementations, namely
KKMChh~\cite{KKMCee,KKMChh,KKMChh19},
Sherpa~\cite{Sherp1,Bothmann:2019yzt}, and Photos~\cite{Was}. In the
latter case, which applies only to QED~FSR, Photos is used within
Powheg+Pythia~\cite{Pow2,Frixione:2007vw,Alioli:2010xd,Alioli:2008gx,Sjostrand:2007gs},
as described in~Section~\ref{sec:samples}. These Monte-Carlo~(MC)
tools provide complete simulated events at LHC~energies, in which the
final state frequently contains several soft/collinear photons near
the final-state leptons from the $Z$-boson decay, in addition to the
single hard large-angle photon emission required by the analysis
selection described in Section~\ref{sec:selection}.
The YFS approach resides on the idea of resumming
the leading soft logarithms to all orders, rather than focusing on the
leading collinear
terms. These soft logarithms are largely independent of the inner
process characteristics and
can be calculated from the external particles and their four-momenta
only. The big advantage
of the YFS formalism is that in addition it allows for a systematic
improvement of this
eikonal approximation, order-by-order in the QED coupling constant
\cite{Mrek1, Mrek2}.
As discussed in
Section~\ref{sec:result}, the measurement results presented in this
paper cover normalised fiducial differential cross-sections and
fiducial integrated cross-sections for \Zllg~decays and a first
measurement of \Zllgg~decays.
 
One often labels the leptons in such final-state configurations as "bare" leptons, a notation adopted throughout this paper, even though the meaning of such a bare lepton is by definition dependent on the tool used to describe multiple-photon emissions in QED FSR. One can also define a "dressed" lepton by combining the bare four-momenta of each lepton with that of QED FSR photons collected close to the lepton within a cone of size $\Delta R$~=~0.1 around it.
 
Since soft/collinear QED emissions are regularised by the lepton mass, the predictions for the kinematic observables describing the final state will differ significantly in some regions of phase space between bare electrons and bare muons when considering the calculations using exponentiation described above. This is shown in the case of predictions from Sherpa 2.2.4~\cite{Bothmann:2019yzt} in~Fig.~\ref{Sherpa_evsmu} for the shape of the normalised distribution for the invariant mass,~\mlg, of the positively charged lepton and the photon and for the angular separation,~\DeltaRlg, between the photon and the closer of the two leptons. For large values of~\mlg, which are close to the edge of the phase space available for \Zllg~decays, and for large values of~$\Delta R_{l\gamma}$, which correspond to specific configurations of \Zllg~decays where the dilepton pair and the photon are back-to-back, Fig.~\ref{Sherpa_evsmu} shows that one expects about~5\% fewer \Zeeg~decays than \Zmmg~decays relative to the bulk of the distribution. It should be noted that most of the difference between electrons and muons observed in these plots for the bulk of the distributions is due to the higher acceptance in~$\eta$ for the muons (see~Table~\ref{cutsTrX}).
 
Figure~\ref{Sherpa_evsmu} also shows that the dressed lepton definition minimises the differences between electrons and muons. The use of dressed leptons also minimises the impact of the QED FSR radiation on the integrated cross-sections. It is often used for exclusive analyses of e.g.~$Z/W$+jet production because it unambiguously assigns the QED FSR photons to the leptons rather than to hadronic jets when performing such exclusive measurements. However, it is not optimal for precise studies of the QED FSR radiation itself.
 
For the purpose of the analysis described in this paper, all results presented in~Section~\ref{sec:result} are shown separately for bare electrons and muons with the exception of the low-statistics measurement of \Zllgg\ final states described in~Section~\ref{sec:Z2l2g}, in which bare electrons and muons are combined. However, all the dressed-lepton results are published in HEPDATA as well. The measured normalised differential distributions of the invariant mass of the negatively charged lepton and the photon are also published in~HEPDATA, together with the charge asymmetry distribution and the full covariance matrix describing the uncertainties in all these measurements with their correlations.

\section{The ATLAS detector}

The ATLAS experiment~\cite{ATLASDET-2008} at the LHC is a multipurpose particle detector
with a forward--backward symmetric cylindrical geometry and a near \(4\pi\) coverage in
solid angle.\footnote{\AtlasCoordFootnote}
It consists of an inner tracking detector surrounded by a thin superconducting solenoid
providing a \SI{2}{\tesla} axial magnetic field, electromagnetic and hadron calorimeters, and a muon spectrometer.
The inner tracking detector covers the pseudorapidity range \(|\eta| < 2.5\).
It consists of silicon pixel, silicon microstrip, and transition radiation tracking detectors.
Lead/liquid-argon (LAr) sampling calorimeters provide electromagnetic (EM) energy measurements
with high granularity.
A steel/scintillator-tile hadronic calorimeter covers the central pseudorapidity range (\(|\eta| < 1.7\)).
The end-cap and forward regions are instrumented with LAr calorimeters
for both the EM and hadronic energy measurements up to \(|\eta| = 4.9\).
The muon spectrometer surrounds the calorimeters and is based on
three large superconducting air-core toroidal magnets with eight coils each.
The field integral of the toroids ranges between \num{2.0} and \SI{6.0}{\tesla\metre}
across most of the detector.
The muon spectrometer includes a system of precision tracking chambers and fast detectors for triggering.
A three-level trigger system is used to select events to be recorded.
The first-level trigger is implemented in hardware and uses a subset of the detector information
to accept events at a rate of at most \SI{75}{\kilo\hertz}.
This is followed by two software-based trigger levels that
together reduce the accepted event rate to \SI{400}{\hertz} on average
depending on the data-taking conditions during 2012.
An extensive software suite~\cite{ATL-SOFT-PUB-2021-001} is used in the reconstruction and analysis of real and simulated data, in detector operations, and in the trigger and data acquisition systems of the experiment.
 
\section{Data and simulated event samples}
\label{sec:samples}
 
The data used in this analysis was collected in~2012 using $pp$~collisions at a centre-of-mass energy of~8~TeV. The corresponding total integrated luminosity  is~20.2$\pm$0.4~fb$^{-1}$. The average number of inelastic $pp$~interactions produced per bunch crossing for this dataset is~20.7.
 
The PowhegBox v1 MC~programme~\cite{Pow2,Frixione:2007vw,Alioli:2010xd,Alioli:2008gx} was used for the simulation of $Z$-boson decays at next-to-leading order~(NLO) in~QCD using the
CTEQ6L1 PDF~set~\cite{CTEQ6}. It was interfaced to Pythia8~\cite{Sjostrand:2007gs} for the modelling of the parton shower, hadronisation, and underlying event, with parameters set according to the AU2~CT10~tune~\cite{ATL-PHYS-PUB-2012-003}. QED FSR is modelled using the exponentiated multi-photon emission provided by Photos~\cite{Davidson:2010ew}. This combination of MC~tools, denoted Powheg+Pythia8+Photos, is used as the main MC~sample to estimate the effects of detector efficiency and resolution and to unfold the data presented in this paper. Initial-state radiation of photons~(QED~ISR) and initial-final-state interference~(IFI) are not included in the simulation. For QED~FSR photons, Photos is configured to include a matrix-element correction for $Z$-boson decay which brings it very close in accuracy (better than~$10^{-3}$) to NNLO~QED~\cite{Was} for the emission of up to two photons.
 
An alternative prediction for $pp \to ll\gamma$ production with up to three additional parton emissions at leading order in the strong coupling is provided by Sherpa~1.4.1~\cite{Sherp1} using the CT10~PDF~set~\cite{CT10}. Sherpa~1.4.1 is configured such that the leading photon is generated at matrix-element level with leading-order accuracy. Both FSR and ISR emissions, but not the small effects expected from~IFI, are included in the simulation. The YFS~\cite{YFS} approach is then used to exponentiate the radiation in a way similar to that done in~Photos. The results obtained by using Sherpa~1.4.1 are compared to those obtained by using Powheg+Pythia8+Photos for a cross-check of the systematics related to the unfolding procedure. A specific version of Sherpa~1.4 (labelled Sherpa~1.4~ME$_{\gamma \gamma}$) containing an exact description at tree level of the production of lepton pairs accompanied by two hard photons  has also been used to compare its predictions to the unfolded \Zllgg\ measurements.
 
For the comparisons between predictions and the unfolded data, additional generator-level predictions for $pp \to ll\gamma$ and $pp \to ll\gamma\gamma$ production have been produced using Sherpa 2.2.4~\cite{Bothmann:2019yzt} and KKMChh~\cite{KKMChh,KKMChh19}. The QED~treatment in Sherpa~2.2.4 is identical to that described above for Sherpa~1.4.1, while the QCD treatment is more advanced~\cite{Sh22}. KKMChh is an event generator for Z boson production and decays, that includes exponentiated multi-photon initial and final state radiation and virtual electroweak corrections. It separates the expected dominant contribution from QED~FSR to the measurements presented in this paper from the negligible contributions from QED~ISR and QED~IFI.
 
To minimise the impact of differences in the QCD aspects on the measurements of QED FSR differential spectra, the MC predictions for which data and~MC are not expected to agree to better than~0.5\% should be reweighted to the data. This procedure was applied for the~\ptZ\ spectrum, which is only predicted at leading order in QCD for the two main MC~samples used in the analysis. Other distributions have also been studied to assess the residual systematic uncertainties from the QCD aspects of the simulation in the measurements. This reweighting procedure is also applied to the MC~predictions at generator level from Sherpa~2.2.4 and from KKMC-hh.
 
Background contributions from top-quark production ($t\bar t$, $Wt$, and $t$-channel single top production) are generated using PowhegBox with the CT10~PDF in conjunction with Pythia6~\cite{Sjostrand:2006za} (v6.427 using the CTEQ6L1~PDF) with the Perugia~2011C tune for parton showering, hadronisation, and underlying event. Background contributions from $Z$+jet and $Z\to \tau\tau$ production have been simulated using Sherpa~1.4.3~\cite{Sherp1} with its  default tune and the CT10~PDFs.
 
The MC~samples were processed through a full ATLAS detector simulation~\cite{SOFT-2010-01}, based on \textsc{Geant4}~\cite{Agostinelli:2002hh}, and reconstructed with the same software as that used for the data. All MC~samples are corrected with data-driven factors to account for small differences in photon and lepton trigger, reconstruction, identification and isolation efficiencies between data and simulation. Additional $p-p$~interactions~(pile-up), occurring in the same and neighbouring bunch crossings are simulated and overlaid at the detector hit level on top of the hard-scattering process from the MC~simulation.

\section{Selection of \Zllg\ events}
\label{sec:selection}
 
A \Zllg\ candidate is formed by selecting the opposite-sign same-flavour dilepton pair with mass closest to~\mZ\ and the highest \pt~photon in the event. No explicit requirements are made on the presence or absence of other activity in the event, such as additional photons or leptons or jets. Only prompt leptons/photons are considered for the signal, meaning that they should not originate from decays of hadrons or $\tau$-leptons. Background events from processes producing fake or non-prompt photons/leptons are suppressed by the lepton/photon identification criteria described below, including in particular isolation requirements on both leptons and on the photon.
 
Event candidates in both data and MC simulation are required to have fired at least one unprescaled single lepton or dilepton trigger, with a lowest \pT~threshold of 24~GeV for single leptons and of~12-13~GeV for dilepton triggers. The single lepton triggers provide a high trigger efficiency with respect to the offline selection, which requires at least one lepton with \pT~$>$25~GeV. The trigger efficiency for \Zllg~events satisfying all the selection criteria described below is about~99\%.
 
\subsection{Photon and lepton selection}
 
Photon and electron candidates are reconstructed~\cite{EGAM-2018-01} from clusters of energy deposits in the electromagnetic calorimeter combined with information from charged tracks reconstructed in the inner detector. Electrons and photons are identified by shower shape and hadronic leakage variables. Photons are required to satisfy all the requirements on shower-shape variables corresponding to the \textit{Tight} photon identification criteria of~Ref.~\cite{PERF-2013-04}. Electron candidates are required to satisfy the \textit{Loose} likelihood requirement of~Ref.~\cite{PERF-2016-01}.
 
Photon (electron) candidates are required to have a transverse momentum above~15~GeV (10~GeV) and a pseudorapidity in the range $|\eta|<2.37$ ($|\eta|<2.47$). For the photon candidates, the transition region between the barrel and end-cap electromagnetic calorimeters, \mbox{$1.37<|\eta|<1.52$}, is excluded.
 
Muon candidates are reconstructed~\cite{PERF-2014-05} from combined tracks in the inner detector and muon spectrometer with a transverse momentum above~10~GeV over the pseudorapidity range $|\eta|<2.5$. Over the range $2.5<|\eta|<2.7$, stand-alone muon candidates in the muon spectrometer are used, based on tracks with reconstructed segments in three spectrometer stations.
 
Electron and muon candidates are required to originate from the appropriate primary vertex (each primary vertex candidate is reconstructed from at least two associated tracks with $\pT > 0.4$~GeV). The significance of the transverse impact parameter, defined as the absolute value of the track transverse impact parameter, $d_{0}$, measured relative to the beam trajectory, divided by its uncertainty, $\sigma_{d_{0}}$, must satisfy $|d_{0}|/\sigma_{d_{0}} < 5$ for electrons and $|d_{0}|/\sigma_{d_{0}} < 3$ for muons. The longitudinal impact parameter is required to satisfy $|z_0| < 10$~mm.
 
The photon, electron, and muon candidates are required to be isolated from other particles using selections based on calorimeter isolation, which is defined as the sum of the calorimeter cell transverse energies, $\sum{E_T}$, inside an isolation cone of $\Delta R$ = 0.2 around the lepton or photon~\cite{PERF-2014-07}. In the case of the calorimeter isolation for photons/electrons, the cells corresponding to the calorimeter cluster itself are excluded from the sum. In addition, the calorimeter isolation is corrected on an event-by-event basis for the energy deposited by the photon or lepton candidate, and, using the method described in Refs.~\cite{Cacciari:2008gn,Cacciari:2009dp, STDM-2010-08}, for the contribution from the underlying event and pile-up.
Each photon/electron (muon) is required to have a normalised calorimeter isolation smaller than~0.3~(0.2). Finally, photons close to leptons are excluded if $\Delta R_{l\gamma} < 0.4$.
 
\subsection{Signal region definition}
 
Candidate \Zllg\ signal events are selected by requiring that they contain at least one pair of opposite-charge, same-flavour leptons and at least one photon. The highest-\pt\ lepton is required to have \pt~$>~25$~\GeV, while the second lepton must have \pt~$>~10$~\GeV. The selected dilepton pair in the event is that which has an invariant mass, $m_{ll\gamma}$, closest to~\mZ. Only 2\% of the events have more than one dilepton pair.
 
\begin{figure}[htb!]
\begin{center}
\includegraphics[width=0.47\textwidth]{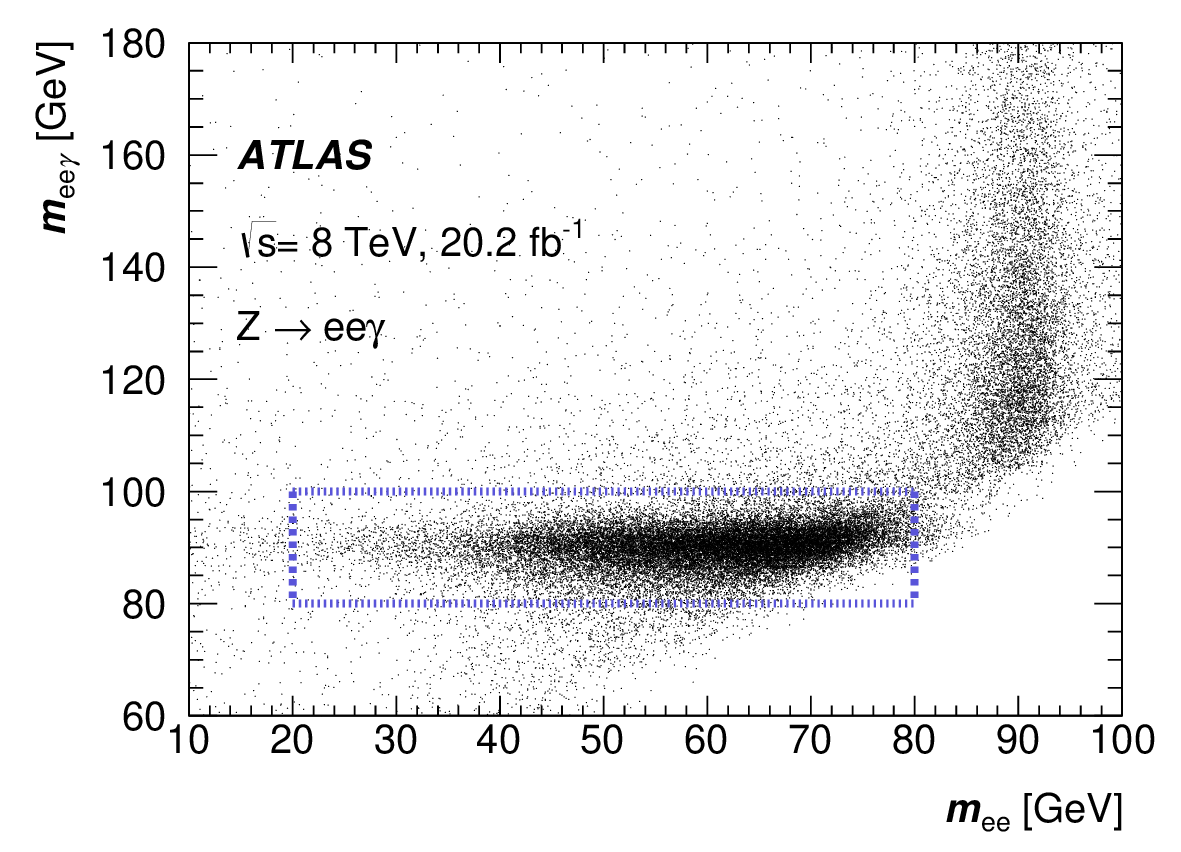}
\hfill
\includegraphics[width=0.47\textwidth]{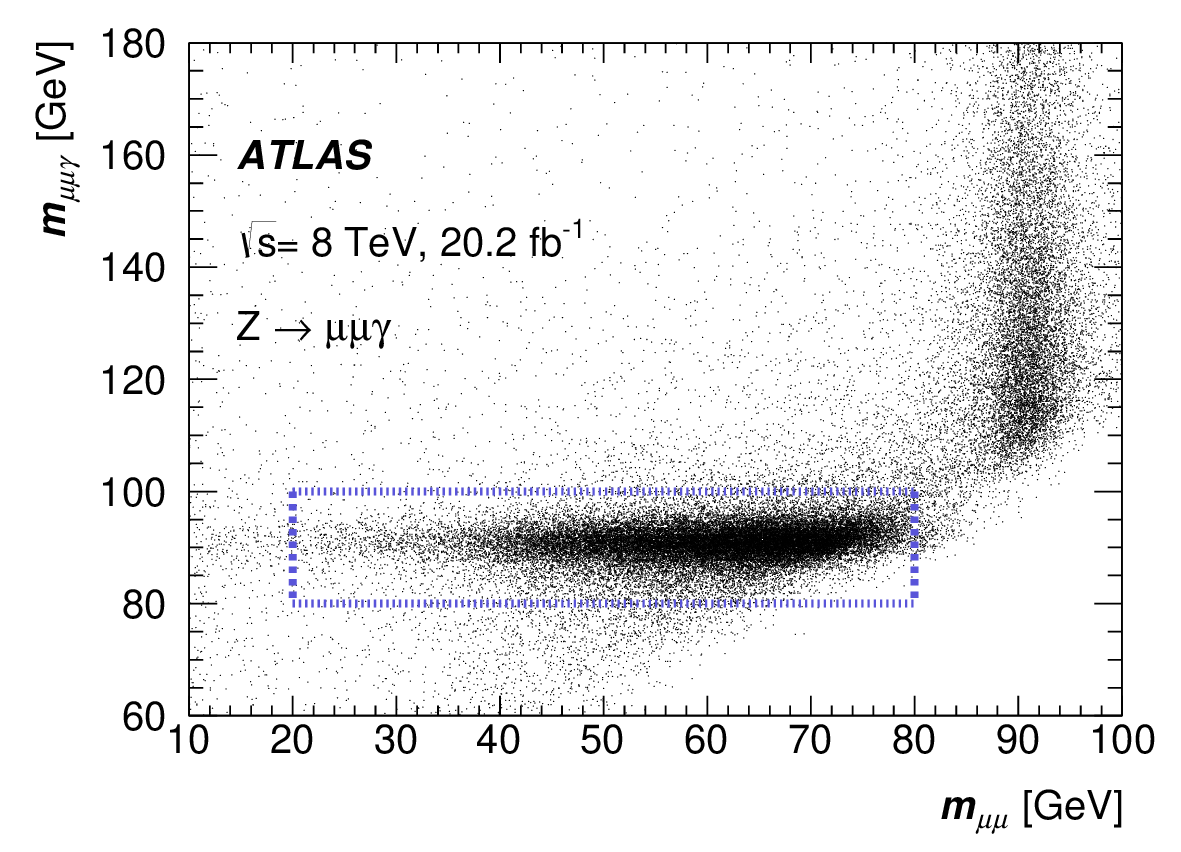}\\
\includegraphics[width=0.47\columnwidth]{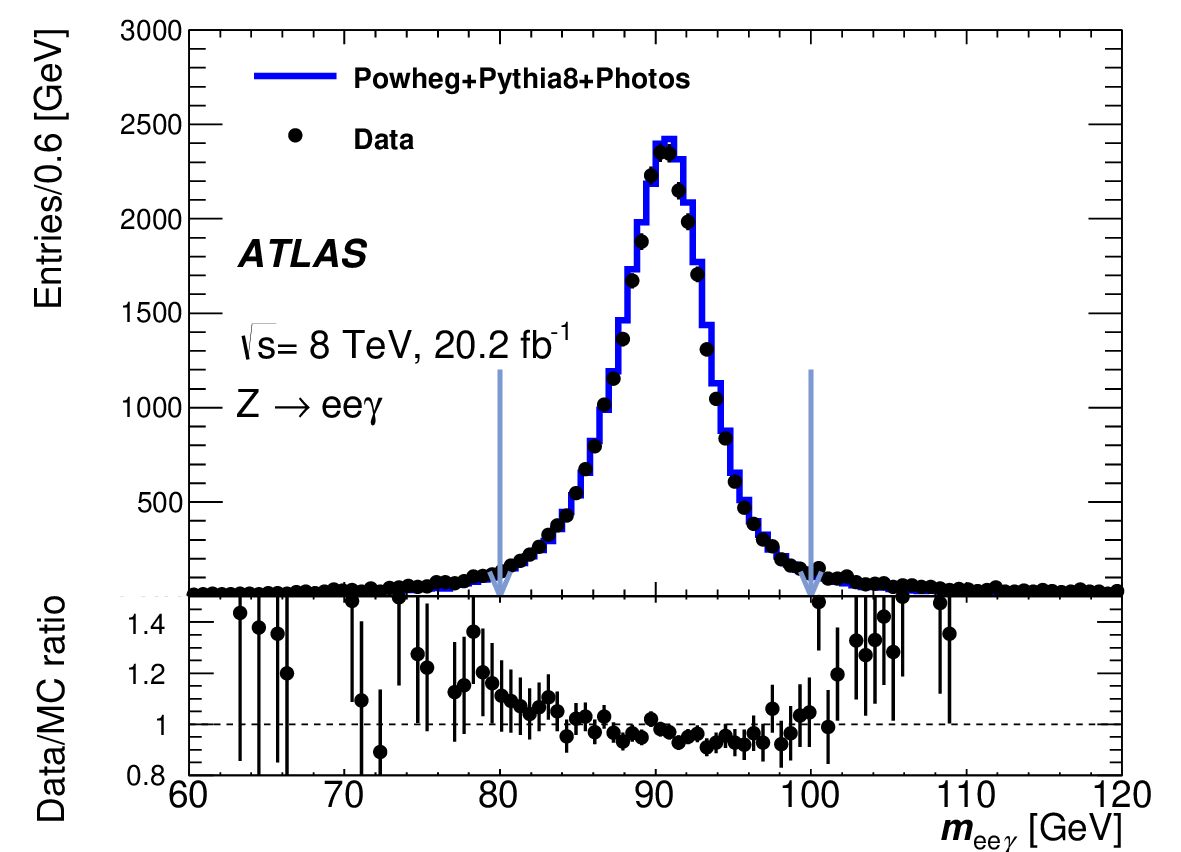}
\hfill
\includegraphics[width=0.47\columnwidth]{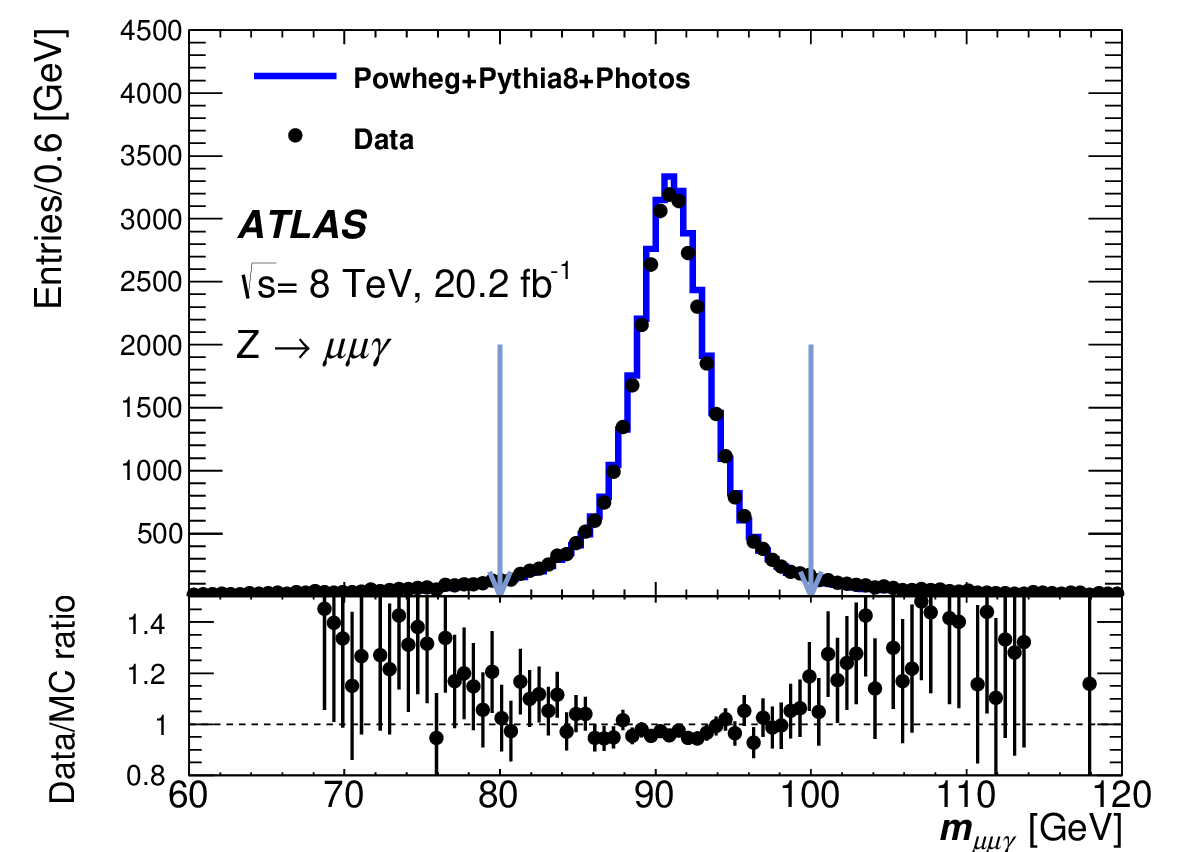}\\
\includegraphics[width=0.47\columnwidth]{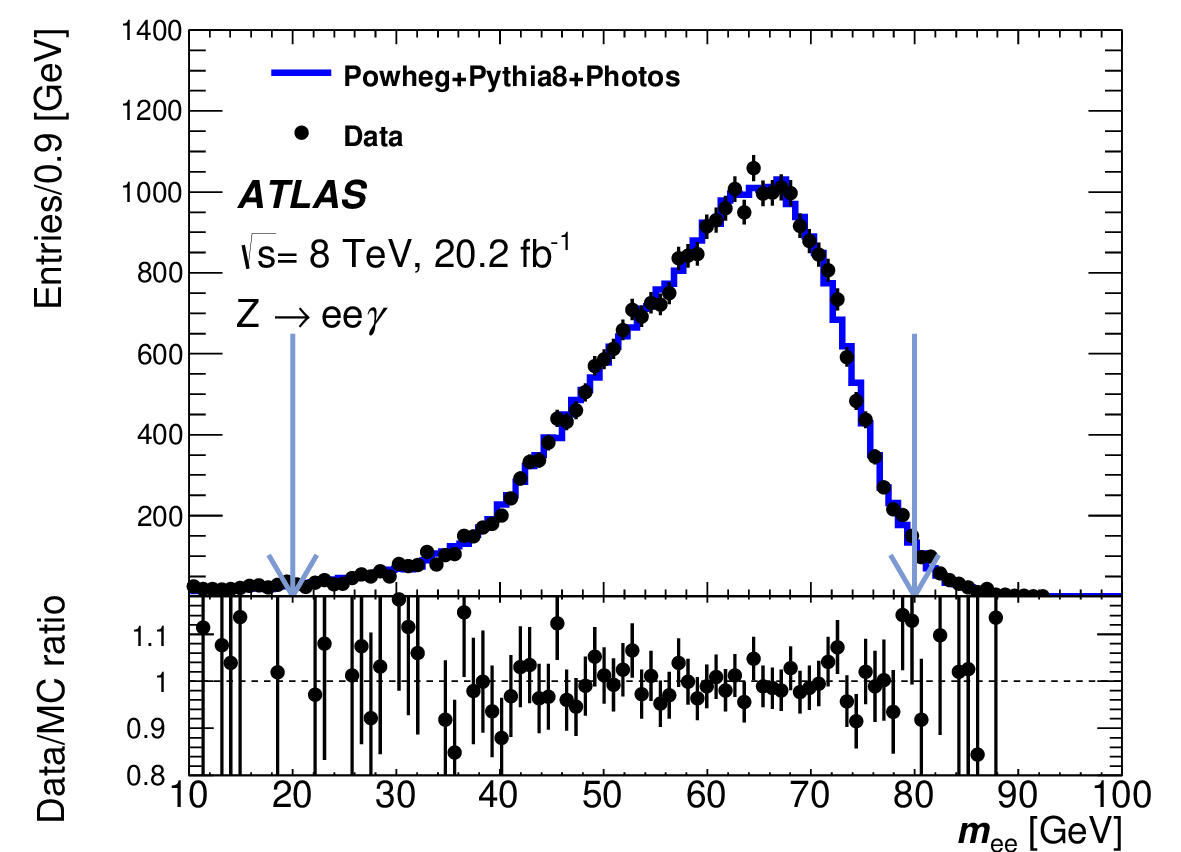}
\hfill
\includegraphics[width=0.47\columnwidth]{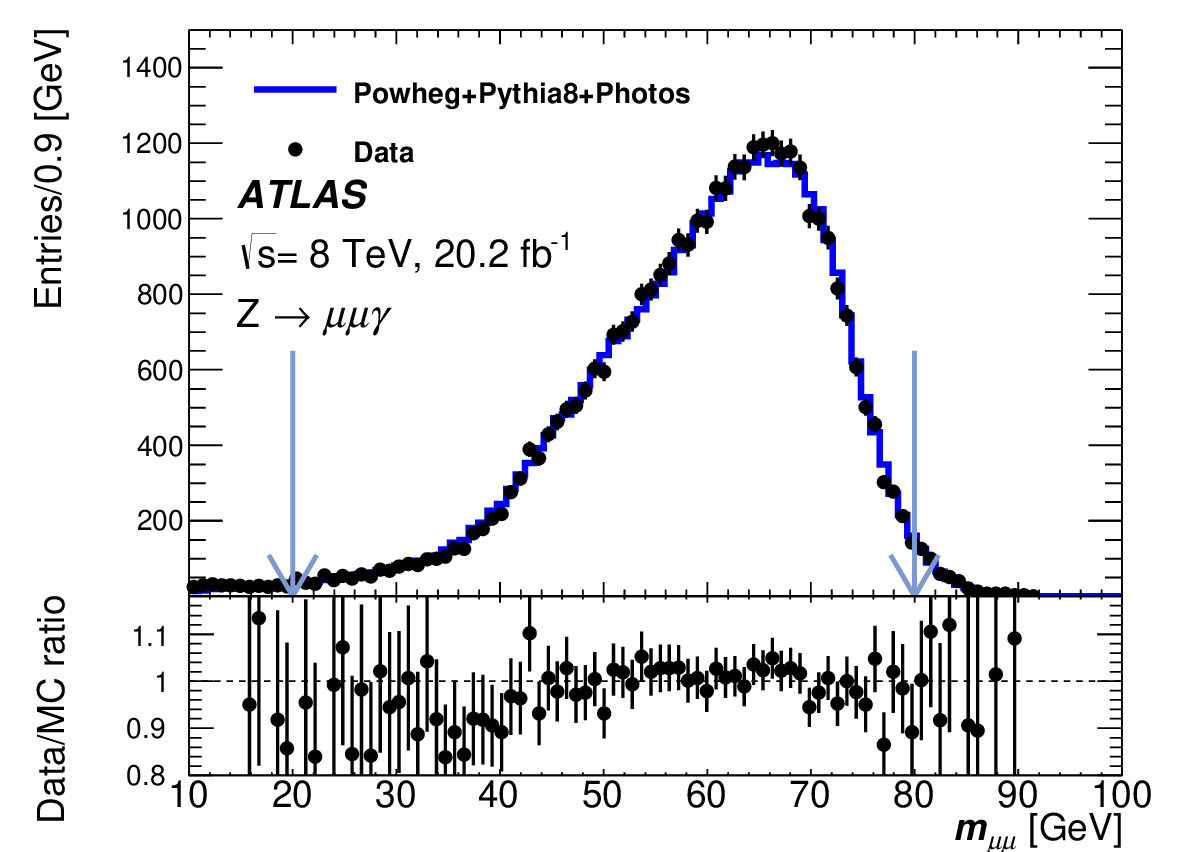}\\

\parbox[t]{1.0\textwidth}{\caption{
For \Zeeg~(left) and \Zmmg~(right) decay candidates in data, distribution of the three-body invariant mass,~$m_{ll\gamma}$, versus the dilepton mass,~$m_{ll}$. The dashed lines indicate the selection criteria applied in the analysis to these two reconstructed invariant masses. These criteria minimise in particular the background from $Z$~bosons produced in association with jets in which one jet is misidentified as a photon. Also shown are the projections of these figures onto the vertical (middle plot with only the selection on~$m_{ll}$ applied) and horizontal (bottom plot with only the selection on~$m_{ll\gamma}$ applied) axes with the corresponding vertical lines indicating the selection cuts described above. The curves show the signal predictions from the nominal Powheg+Pythia8+Photos~MC normalised to the data.
\label{mass0}
}}
 
\end{center}
\end{figure}

To further improve the purity of the QED FSR sample, the invariant mass of the dilepton pair, $m_{ll}$, is required to be between 20 and 80 GeV, while the three-body invariant mass, $m_{ll\gamma}$, is required to be between 80 and 100 GeV. These selections mainly suppress the background from $Z$-boson+jet production with a jet misidentified as a photon, which can be clearly seen in~Fig.~\ref{mass0} outside the signal region. For similar reasons, these selections also reduce considerably the background from QED~ISR. The integrated fiducial cross-section results are also extracted for a stricter selection on the invariant mass of the dilepton, $m_{ll} >~$45~GeV.
 
\begin{figure}[htb!]
\begin{center}
\includegraphics[width=0.48\columnwidth]{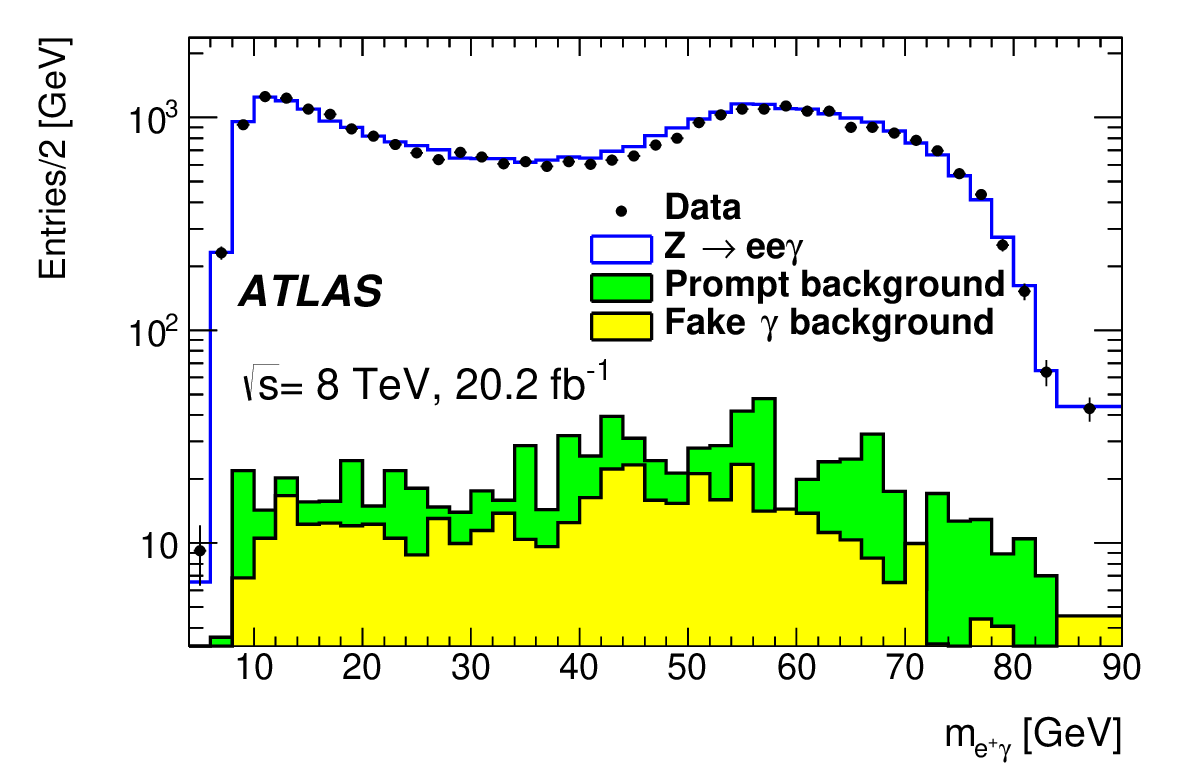}
\hfill
\includegraphics[width=0.48\columnwidth]{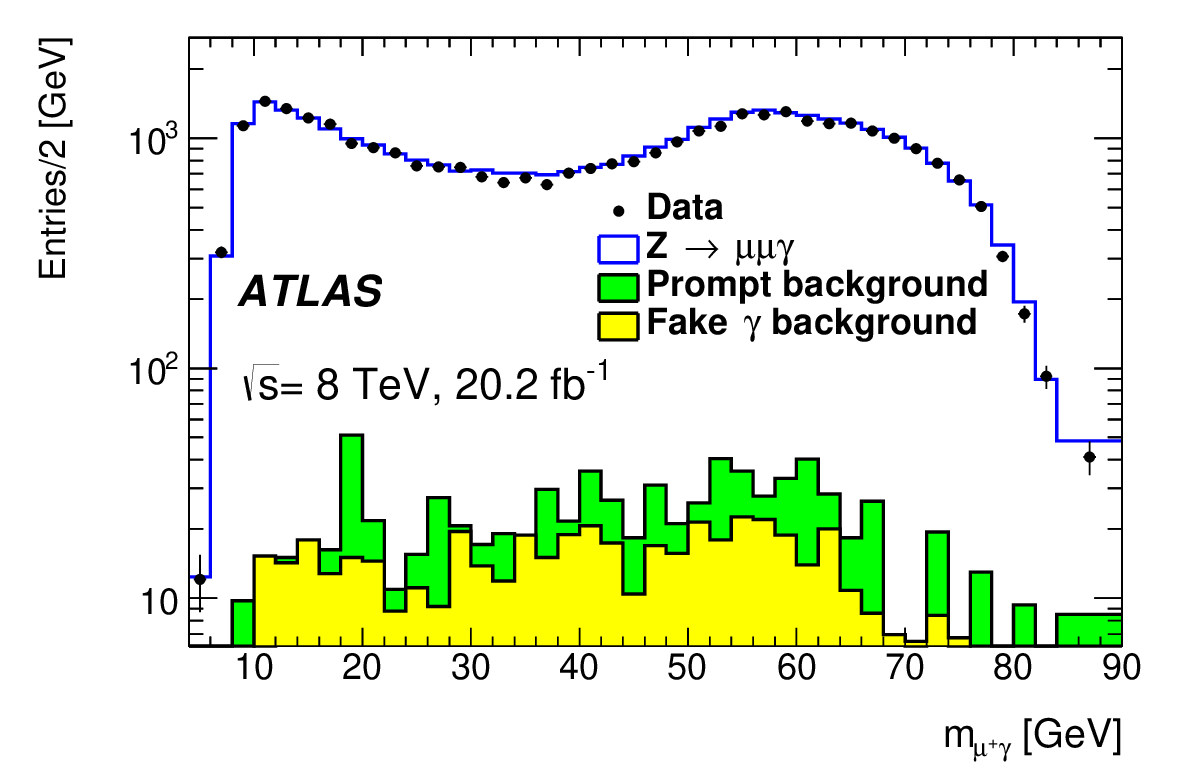}\\
\includegraphics[width=0.48\columnwidth]{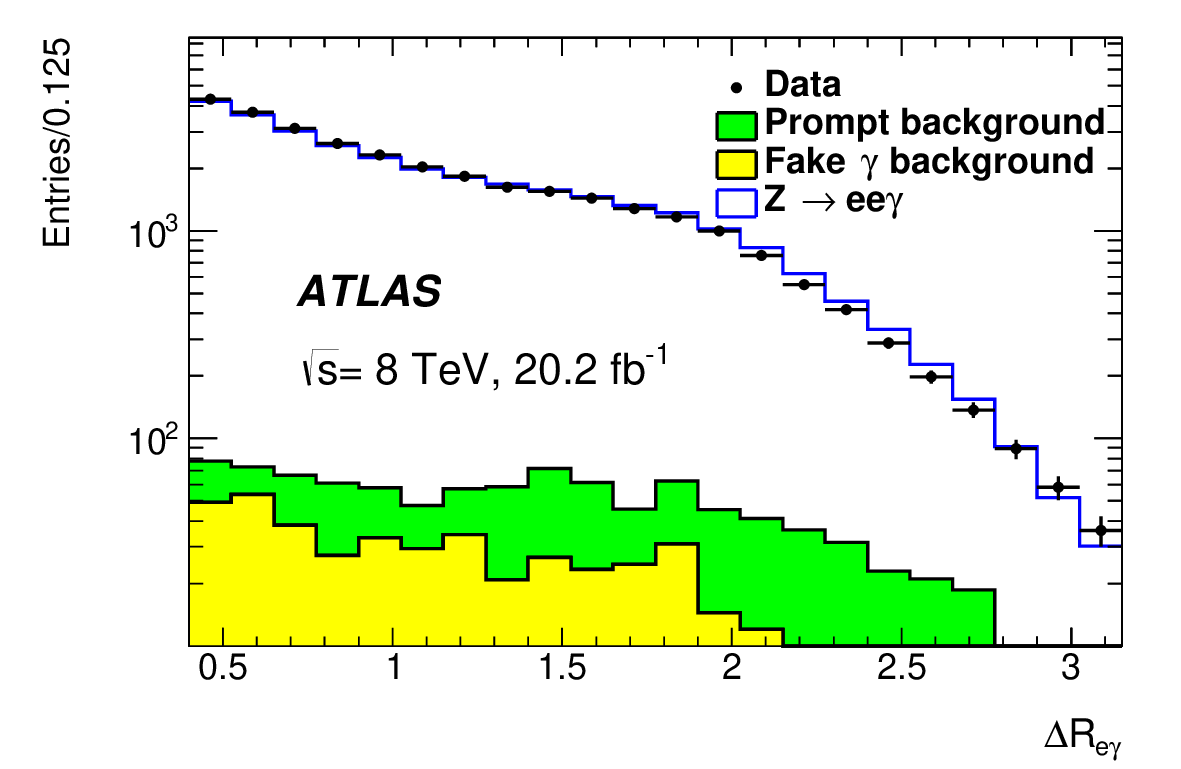}
\hfill
\includegraphics[width=0.48\columnwidth]{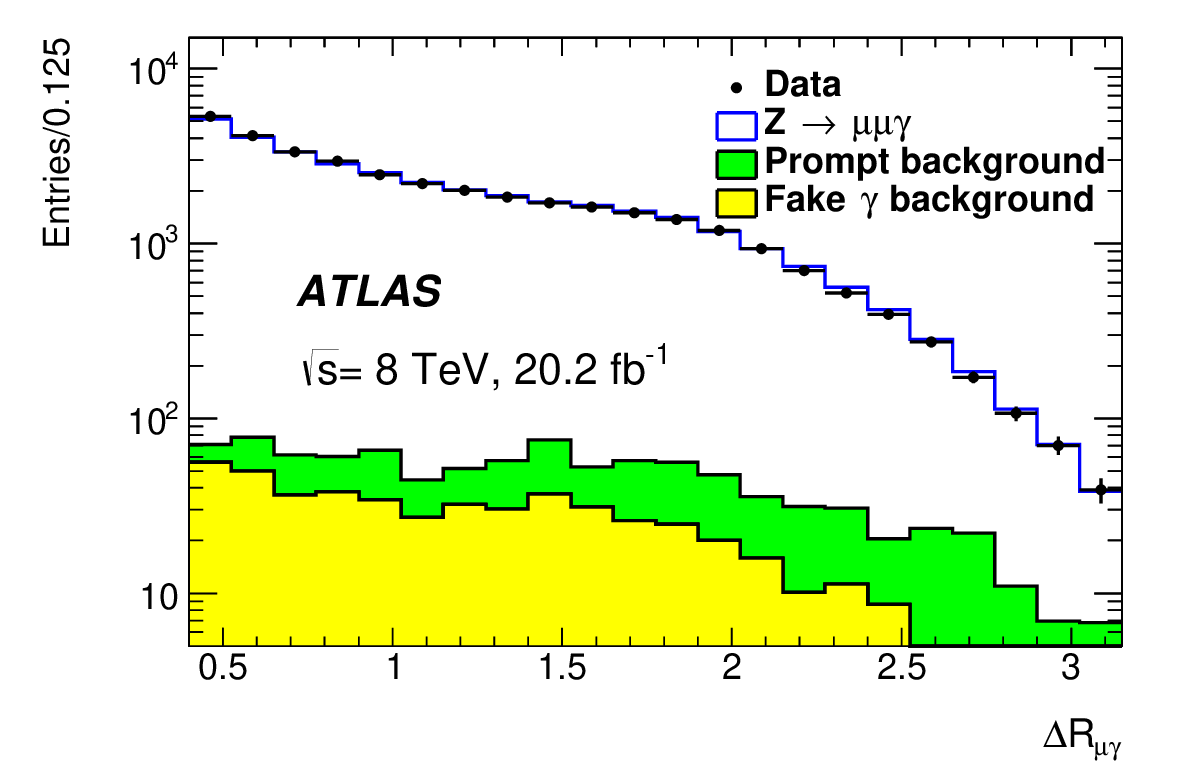}\\
\includegraphics[width=0.48\columnwidth]{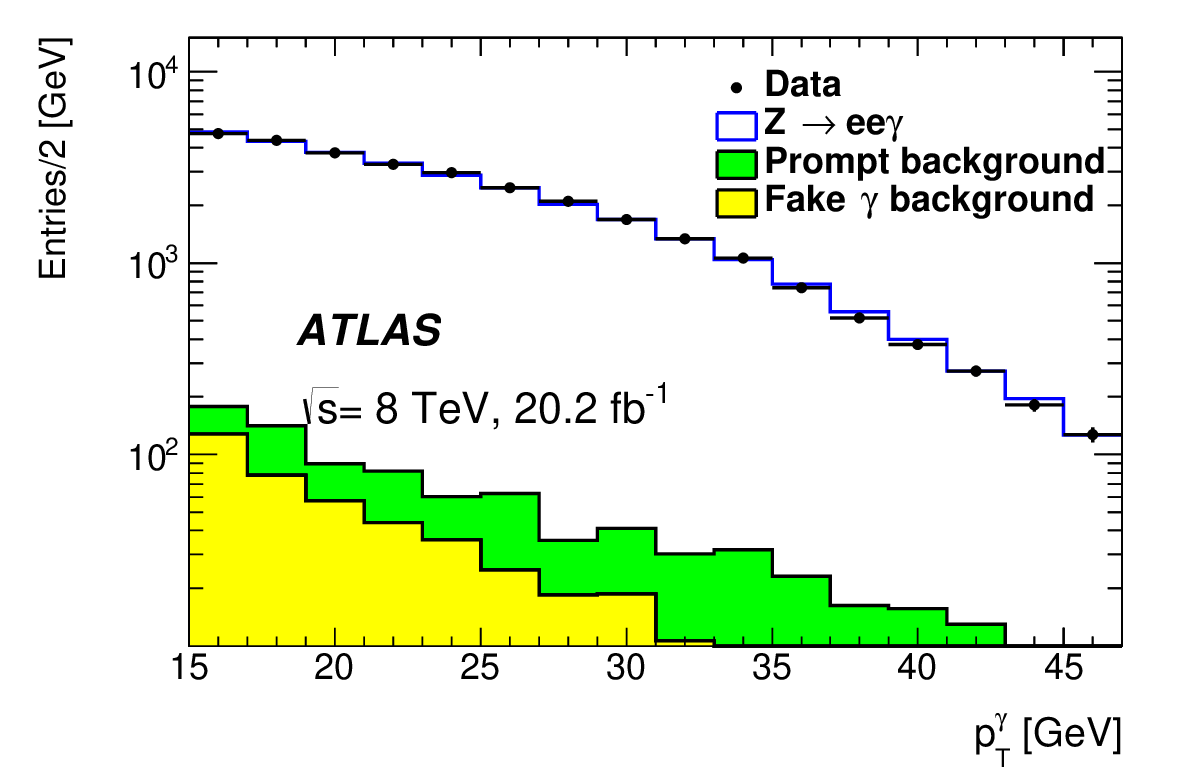}
\hfill
\includegraphics[width=0.48\columnwidth]{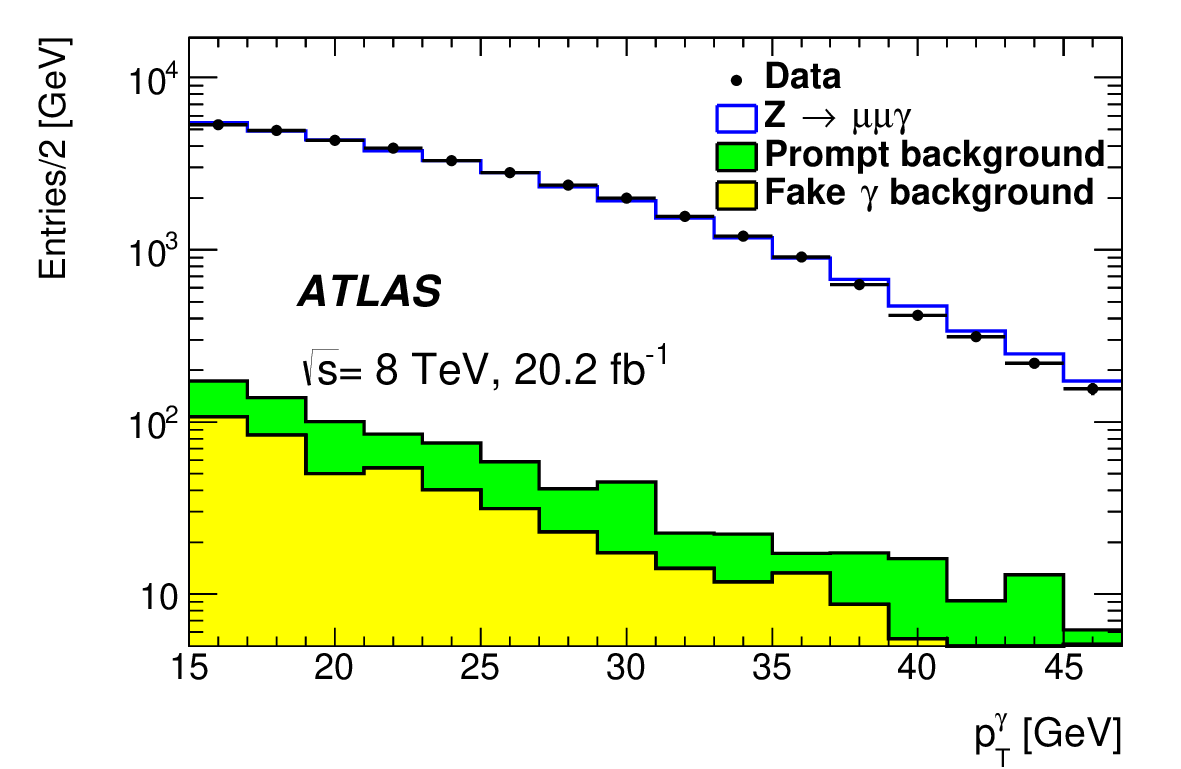}\\
\caption{
For electrons (left) and muons (right), measured differential rates for~\mlg~(top), \DeltaRlg~(middle), and \ptg~(bottom). The data are shown compared to the sum of the expected contributions from the \Zllg~signal, as predicted from Powheg+Pythia8+Photos, and of the prompt and fake~$\gamma$~backgrounds.
\label{fonX}
}
\end{center}
\end{figure}

The criteria described above were applied to the data, leading to~30571 (resp.~34948) events selected in the \Zeeg\ (resp.~\Zmmg) channel (see Table~\ref{Fon1xm} and Section~\ref{sec:backgrounds} for more details). The reconstructed differential distributions are shown in~Fig.~\ref{fonX} for the three observables of interest, the invariant mass of the positively charged lepton and the photon,~\mlg, the angular separation between the photon and the closer of the two leptons,~\DeltaRlg, and the transverse momentum of the photon,~\ptg.
 
\section{Background estimation}
\label{sec:backgrounds}
 
The dominant sources of background to the \Zllg~signal originate from
top-quark pair production and decay with two high-\pt\ prompt leptons
and one prompt photon in the final state and from $Z$+jet production
with two high-\pt\ prompt leptons and one jet misidentified as a
photon in the final state. Other smaller sources of background are
also considered,  such as $WZ\gamma$~production,
$\tau\tau\gamma$~production with two leptonic decays of
the~$\tau$-lepton, but also backgrounds with one fake or non-prompt
lepton such as $W\gamma$+jet production and backgrounds from $WZ$
production with the electron from the $W$-boson decay misidentified as
a photon.
The sum of all sources contributing to events containing two prompt leptons and one prompt photon is labelled "prompt background" and is predicted by simulation, while the sum of all other sources is labelled "fake~$\gamma$~background" and is estimated from data as described below. The respective contributions of these prompt and fake~$\gamma$ background sources to the data are listed separately with their total uncertainties for the dielectron and dimuon case in~Table~\ref{Fon1xm}.

\begin{table}[htb!]
\begin{center}
\caption{ \label{Fon1xm}
Total observed rates of data selected for the measurements in the \Zeeg\ and \Zmmg\ channels. Also shown are the expected contributions with their total uncertainties from the two main sources of background, the prompt background containing two leptons and one photon, predominantly from top-quark pair production, and the fake~$\gamma$~background containing two leptons and one jet misidentified as a photon, predominantly from $Z$+jet production (see text). The last row shows the expected signal with its total uncertainty (see Section~\ref{sec:intXsec}).
}
\vspace{1cm}
\begin{tabular}{|l|c|c|}
\hline
Channel & \Zeeg & \Zmmg \\  \hline
Data & 30571 &  34948 \\ \hline
Prompt background & 360 $\pm$ 40 &  290 $\pm$ 50  \\
Fake $\gamma$ background & 450 $\pm$ 90 &  500 $\pm$ 90 \\
\hline
Total background & 810 $\pm$ 100 &  790 $\pm$ 100 \\
\hline
\Zllg\ expected signal & 28990 $\pm$ 990 &  34530 $\pm$ 1100 \\
\hline
\end{tabular}
\end{center}
\end{table}

The overall background contribution to the signal is expected to be
small, as can be seen from the small tails outside the pole region in
the three-body invariant mass,~$m_{ll\gamma}$, distribution
of~Fig.~\ref{mass0}. A simple extrapolation from the events with
large~$m_{ll\gamma}$ to the signal region leads to an expectation at
the percent level. The nominal method used for the background
estimation in this analysis is an unbinned maximum likelihood fit of
the $m_{ll\gamma}$~distribution over a wide
range,~$60<m_{ll\gamma}<120$ GeV, using simulation for the prompt
background sources (shape and normalisation) and also for the shape of
the fake~$\gamma$ background. This fit is performed for each bin of
the differential cross-section measurements.
 
A cross-check of the fake~$\gamma$ background estimate is performed with a data-driven two-dimensional sideband method using an inverted isolation selection for the photon candidate on one side and requiring that $100 < m_{ll\gamma} < 150$~GeV on the other. This latter selection enhances the fake~$\gamma$ background (see~Fig.~\ref{mass0}~(top)). The difference between the two methods is used as an estimate of the systematic uncertainty in the fake~$\gamma$ background estimate. As shown in~Table~\ref{Fon1xm}, the overall background is at the level of~2-3\% of the expected signal with a total relative uncertainty of~13\%.

\section{Correction for detector effects}
\label{sec:unf}
 
The reconstructed differential distributions presented in~Fig.~\ref{fonX} are corrected for detector effects and bin-to-bin
migrations using an iterative Bayesian unfolding method applied to events that pass the detector-level selections. The method is applied with three iterations implemented in the RooUnfold framework~\cite{UnfB}. The resulting unfolded distributions for the \Zeeg\ and \Zmmg\ signal are obtained at both bare-lepton and dressed-lepton level.
 
The particle-level selections shown in~Table~\ref{cutsTrX} are chosen to be as close as possible to the selections described in~Section~\ref{sec:selection}. The leptons and the photon are required to be prompt, meaning that they should not originate from decays of hadrons or $\tau$-leptons. The small difference in pseudorapidity ranges for electrons and muons is retained at particle level since the electron and muon channels are not combined in this analysis (except for the low statistics \Zllgg~sample described in~Section~\ref{sec:Z2l2g}).
 
The response matrices, which connect the distributions at reconstruction and particle level, as well as the efficiency correction factors are derived using the Powheg+Pythia+Photos signal MC~sample. The reconstruction, trigger and isolation efficiencies as well as the photon/lepton momentum scale and resolution in the MC simulation are corrected to match those determined in data. The transverse momentum distribution of the simulated \Zllg~events was reweighted to that observed in data to minimise the only potentially significant bias from QCD modelling in the simulation (see Section~\ref{sec:sys} for more details). Since the charged leptons and the photon are precisely measured compared to the bin sizes chosen for the differential distributions, the systematic uncertainties related to the unfolding method were found to be below~0.1\%.

\begin{table}[htb!]
\begin{center}
\caption{\label{cutsTrX} Fiducial volume selections for \Zllg\ events}
\vspace{3mm}
\begin{tabular}{|c|}
\hline
Photon with \pt~$> 15$ GeV \\
$|\eta_{\gamma}|<2.37$ excluding $1.37<|\eta_{\gamma}|<1.52$ \\
Leptons with \pt~$> 25, 10$ GeV\\
$|\eta_{\mu}| < 2.7$ \\
$|\eta_e| < 2.47$ \\
At least one photon + one pair of same-flavour opposite-sign leptons \\
\DeltaRlg~$> 0.4$ \\
$20 < m_{ll} < 80$ GeV \\
$80 < m_{ll\gamma} < 100$ GeV \\
\hline
\end{tabular}
\end{center}
\end{table}

\section{Systematic uncertainties}
\label{sec:sys}
 
\begin{figure}[htb!]
\begin{center}
\includegraphics[width=0.49\columnwidth]{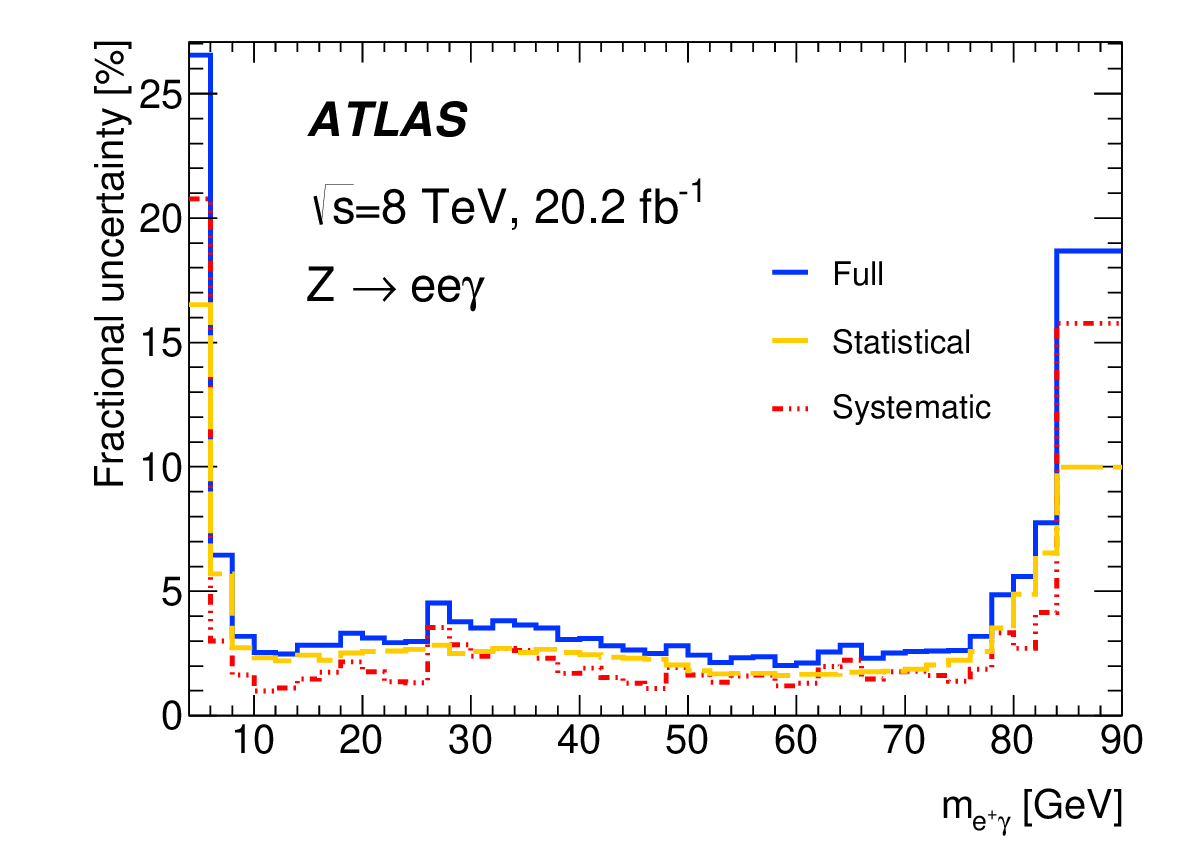}
\hfill
\includegraphics[width=0.49\columnwidth]{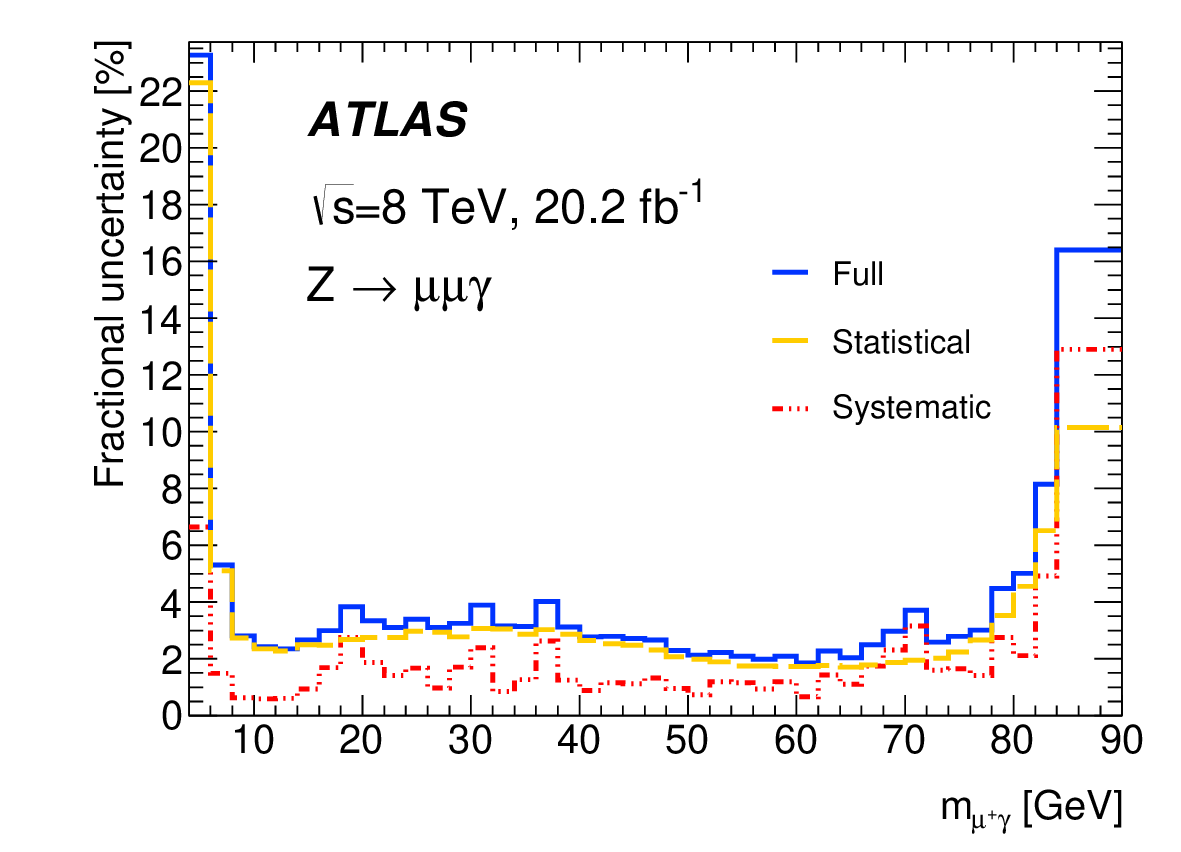}\\
\includegraphics[width=0.49\columnwidth]{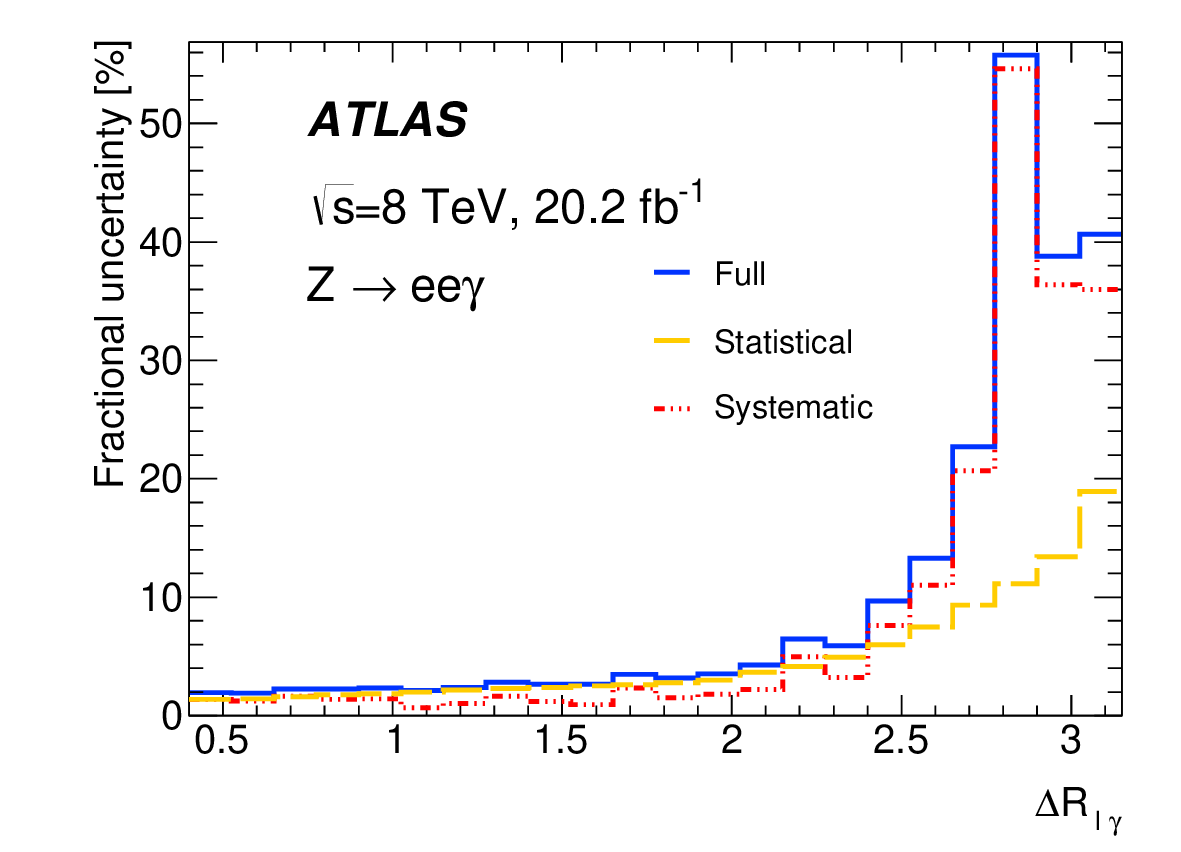}
\hfill
\includegraphics[width=0.49\columnwidth]{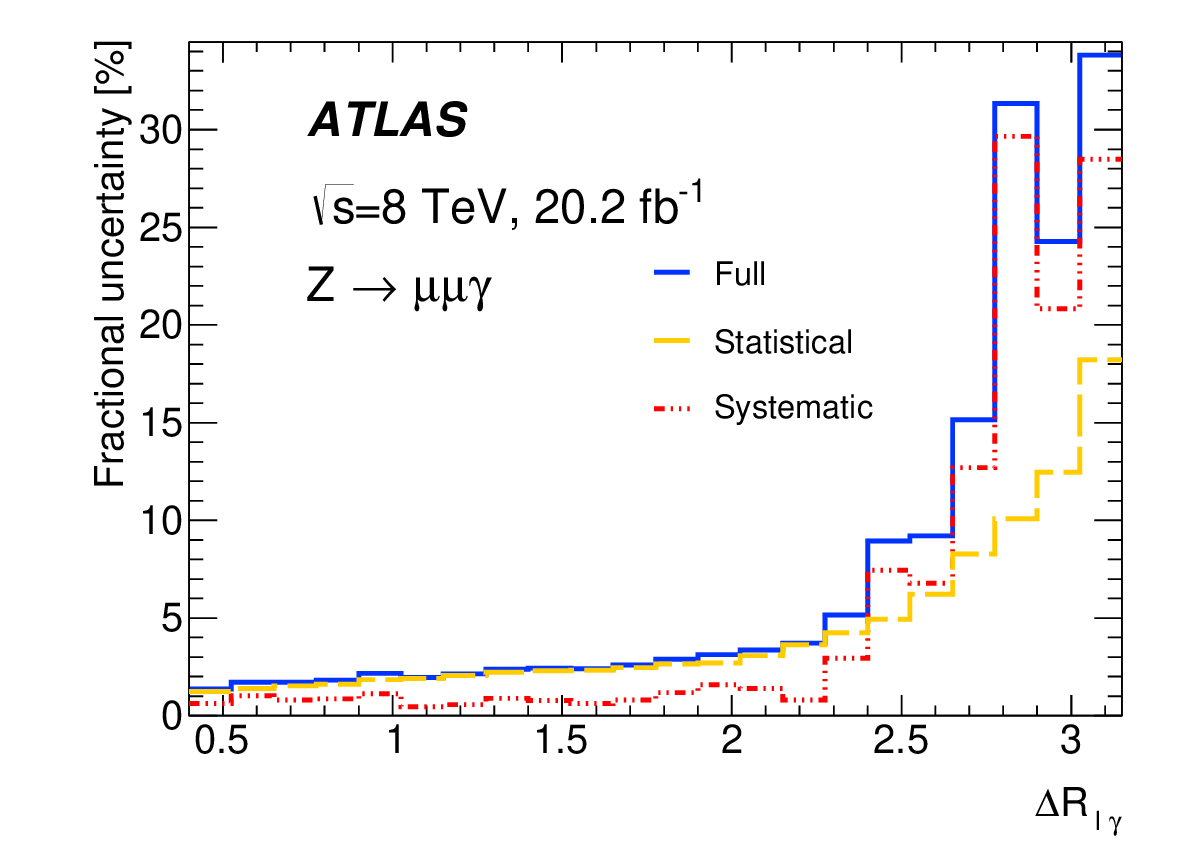}\\
\includegraphics[width=0.49\columnwidth]{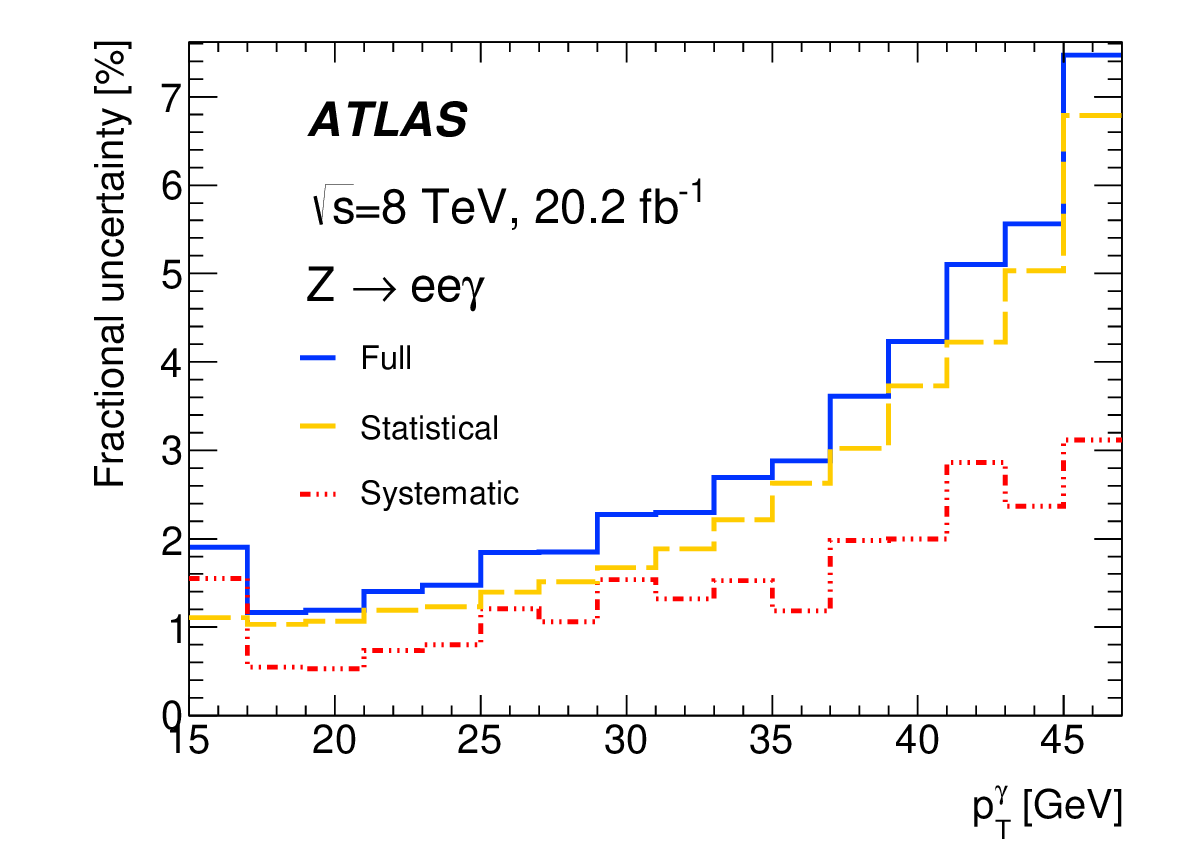}
\hfill
\includegraphics[width=0.49\columnwidth]{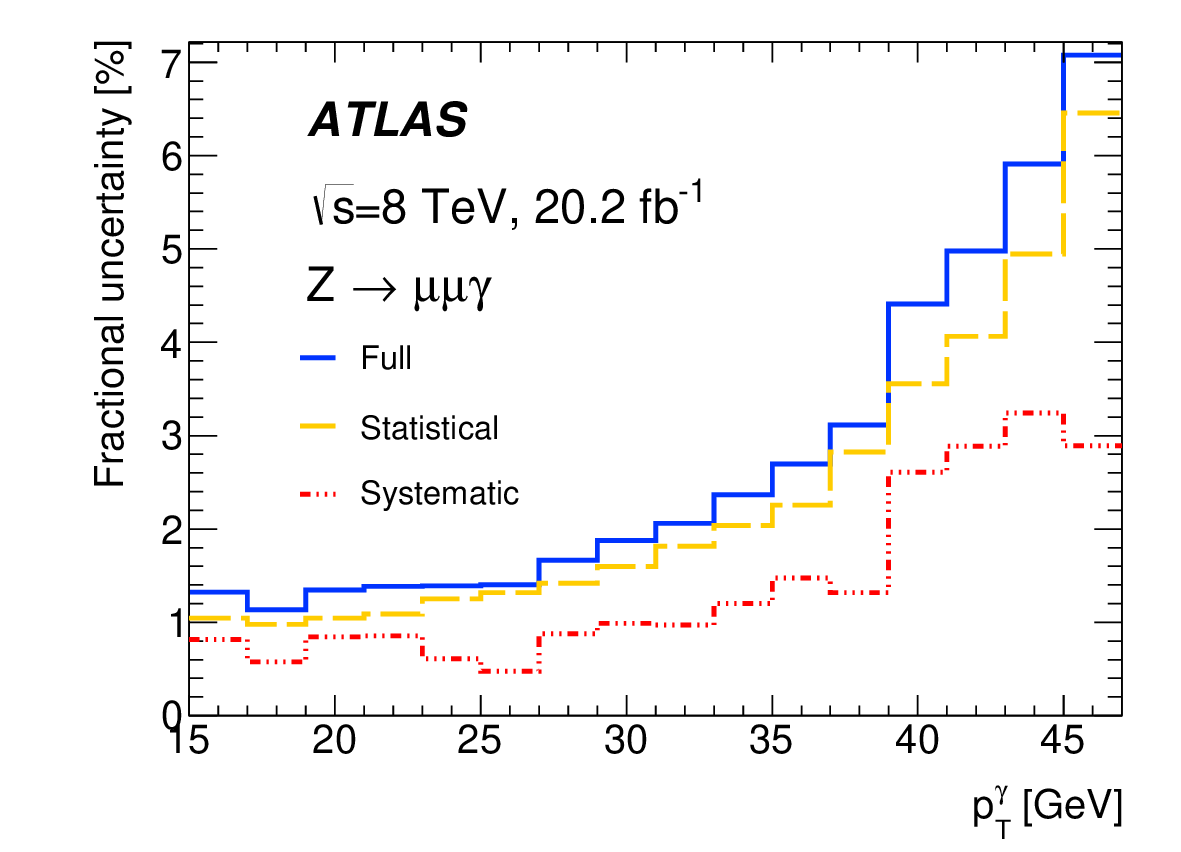}\\
\caption{Breakdown of relative uncertainties for the normalised unfolded differential cross-sections as a function of the observables of interest, \mlg~(top), \DeltaRlg~(middle), and~\ptg~(bottom), shown separately for the \Zeeg~(left) and \Zmmg~(right) channels.
\label{errMlg0}}
\end{center}
\end{figure}

This section describes the systematic uncertainties in the measurements of the normalised differential fiducial cross-sections for the observables presented in~Fig.~\ref{fonX}. These systematic uncertainties are grouped according to their source, and their typical relative values over most of the kinematic range of the measured observables are listed in~Table~\ref{SysTotX}, while the relative variations of the overall systematic uncertainty as a function of the measured observables are shown in~Fig.~\ref{errMlg0}, together with the variations of the statistical uncertainty in the data and of the total uncertainty in the measurements. The statistical uncertainty in the data is the dominant uncertainty over most of the measurement bins of all three observables.
 
The experimental systematic uncertainties in measured observables involving only leptons and photons are expected to be small, and particularly so for normalised distributions for which only the shape variations in these uncertainties remain. This can indeed be seen from~Table~\ref{SysTotX} which shows contributions of at most a few per mille from the lepton/photon experimental systematic uncertainty sources. However, Fig.~\ref{errMlg0} shows that the overall systematic uncertainty is the dominant contribution at the edge of the phase space for the~\mlg\ and~\DeltaRlg\ distributions. This arises from the very limited number of simulated events in these regions which inflates the overall systematic uncertainty even though it is statistical in nature.

\begin{table}[htb]
\begin{center}
\caption{ \label{SysTotX} Breakdown of systematic uncertainties for the normalised differential fiducial cross-sections. The values shown are typical over most of the kinematic range of the unfolded observables. }
\vspace{5mm}
\begin{tabular}{|l|c|c|}
\hline
Uncertainty source & \Zeeg\ channel & \Zmmg\ channel \\ \hline
Experimental &  &  \\ \hline
Energy/momentum scale and resolution & 0.2\% & 0.2\% \\
Efficiency & 0.3\% & 0.3\% \\
Unfolding & $<$ 0.1\% & $<$ 0.1\%\\
Background subtraction & 0.3\% & 0.3\% \\
\hline
Theory  & & \\ \hline
PDF & $<$ 0.1\% & $<$ 0.1\%\\
QCD scale variations & 0.1\% & 0.1\%\\
QCD modelling & 0.3\% & 0.3\%\\
\hline
Total & 0.6\% & 0.6 \%\\
\hline
\end{tabular}
\end{center}
\end{table}

Over most of the phase space of interest, one expects that the shapes of the electron and muon channels are identical, so the $e-\mu$ difference has been used as an additional cross-check of the experimental systematic uncertainties. The weighted mean of the $e-\mu$ data/MC double ratio difference was found to be
$-0.0011 \pm 0.0085$~(stat)~$\pm$~0.00007~(syst). The systematic uncertainty is reduced here purely to the contribution of the experimental systematic uncertainties which are uncorrelated between electrons and muons. The $e-\mu$ difference was thus found to be consistent with its statistical uncertainty, and no systematic trend was observed as a function of the measured differential distributions.

Table~\ref{SysTotX} shows that the theoretical uncertainties from QCD
scale variations and PDFs, related to the acceptance of the
photon/lepton fiducial cuts, are very small, at the per mille level,
as expected. However, any QCD-related mis-modelling of $Z$-boson
production not covered by these uncertainties may have a significant
and direct impact on the QED-related observables of interest. As
mentioned above, the simulated events were reweighted to reproduce the
observed transverse momentum spectrum of \Zllg~decays. This has
potentially  a direct impact on the photon transverse momentum
spectrum, but it also affects at the percent level certain regions of
phase space for the \mlg\ and \DeltaRlg~distributions (the most
significant impact has been observed in the case of the Sherpa 1.4
predictions). After this reweighting procedure, the residual QCD
modelling systematic uncertainties were found to be very small, at the
few per mille level. This was further verified by also reweighting to
data the other kinematic observables related to $Z$-boson production,
namely the $Z$-boson rapidity and the angular
coefficients~\cite{STDM-2014-10} related to the lepton kinematics in
the rest-frame of the $Z$-boson.
As a final cross-check, the Sherpa
1.4 predictions for the reconstructed observables, after reweighting
to the observed transverse momentum spectrum of \Zllg~decays, were
unfolded using the nominal Powheg+Pythia+Photos~MC. The resulting
systematic uncertainty, as derived from the difference between the
generator-level Sherpa 1.4 prediction and that obtained after this
unfolding cross-check, was found to be negligible,~i.e.~below the per
mille level.

\section{Results for \Zllg~process}
\label{sec:result}
 
\subsection{Differential cross-sections}
 
The measurement results are presented here as unfolded normalised
differential cross-sections for the three observables of interest, namely the invariant mass of the positively charged lepton and the photon,~\mlg, the angular distance,~\DeltaRlg, between the photon and the closer of the two leptons, and the transverse momentum,~\ptg, of the photon. These differential cross-sections are normalised to the total measured fiducial cross-section to minimise the experimental uncertainties and also to remove potentially dominant QCD-related theoretical uncertainties when comparing to predictions. The results below are shown separately for the electron and muon channels in the case of bare leptons. This maximises the sensitivity to expected small differences in the distributions between the two lepton flavours (see~Figure~\ref{Sherpa_evsmu}).
 
\begin{figure}[htb!]
\begin{center}
\includegraphics[width=0.49\columnwidth]{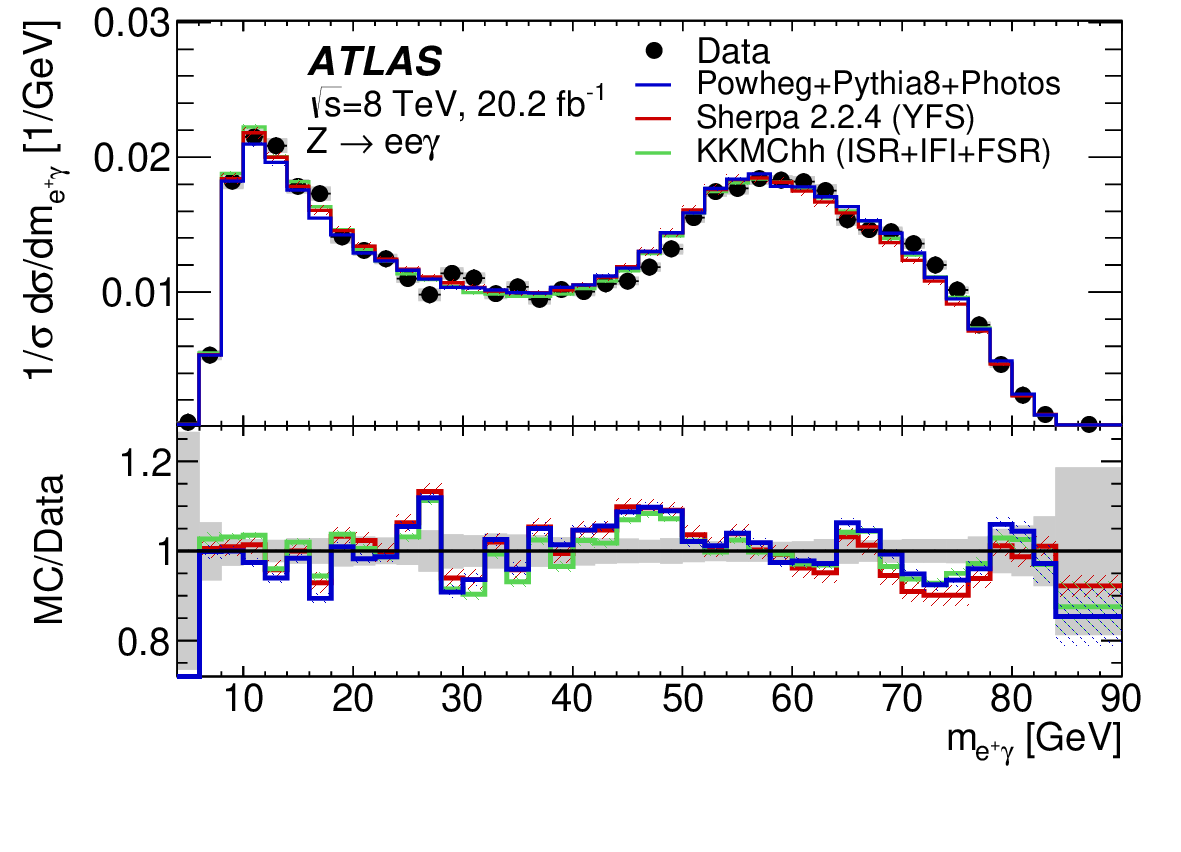}
\hfill
\includegraphics[width=0.49\columnwidth]{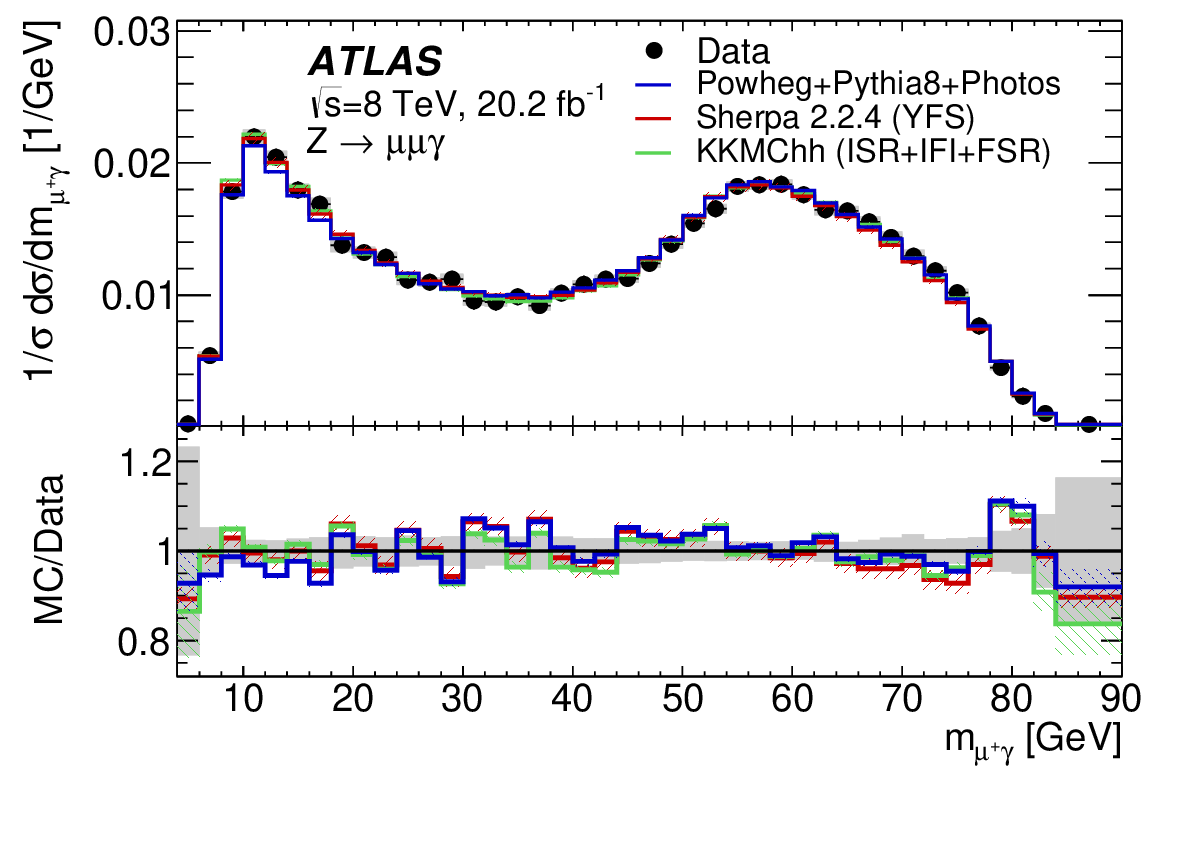}
\includegraphics[width=0.49\columnwidth]{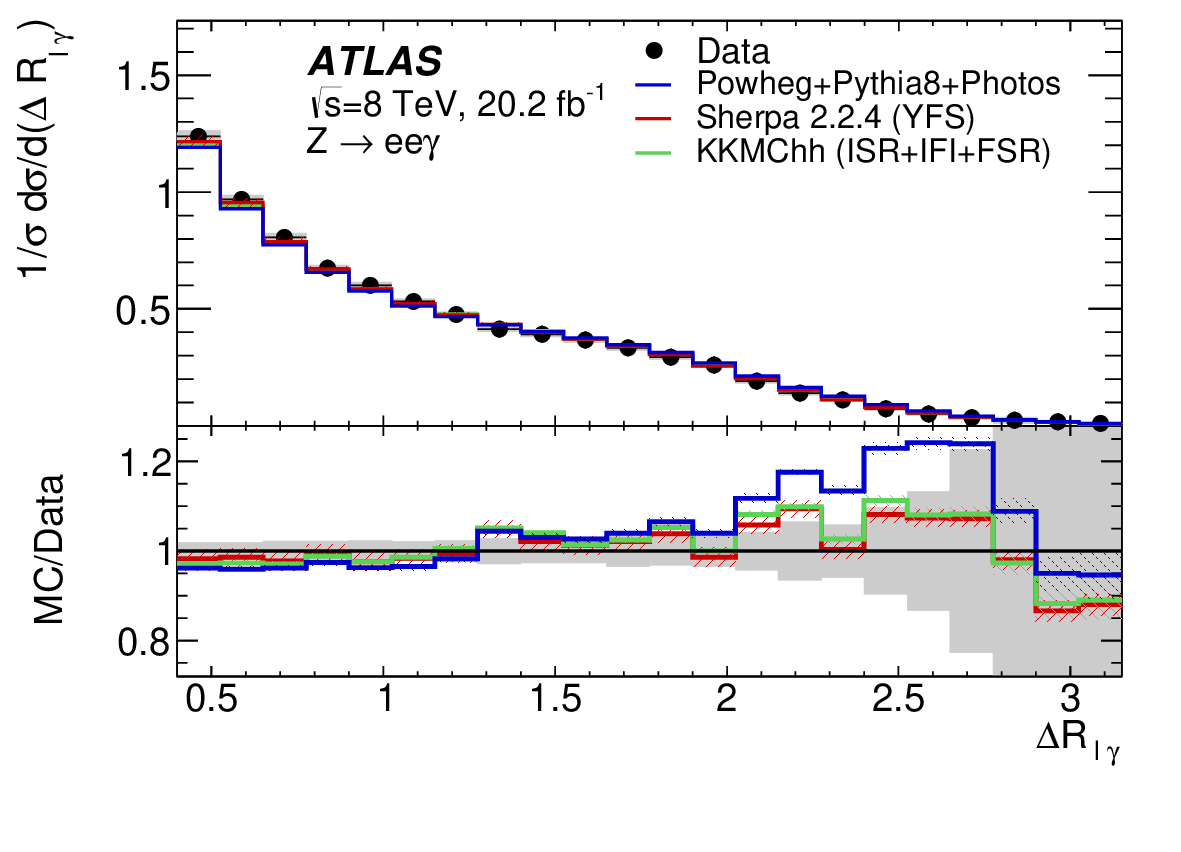}
\hfill
\includegraphics[width=0.49\columnwidth]{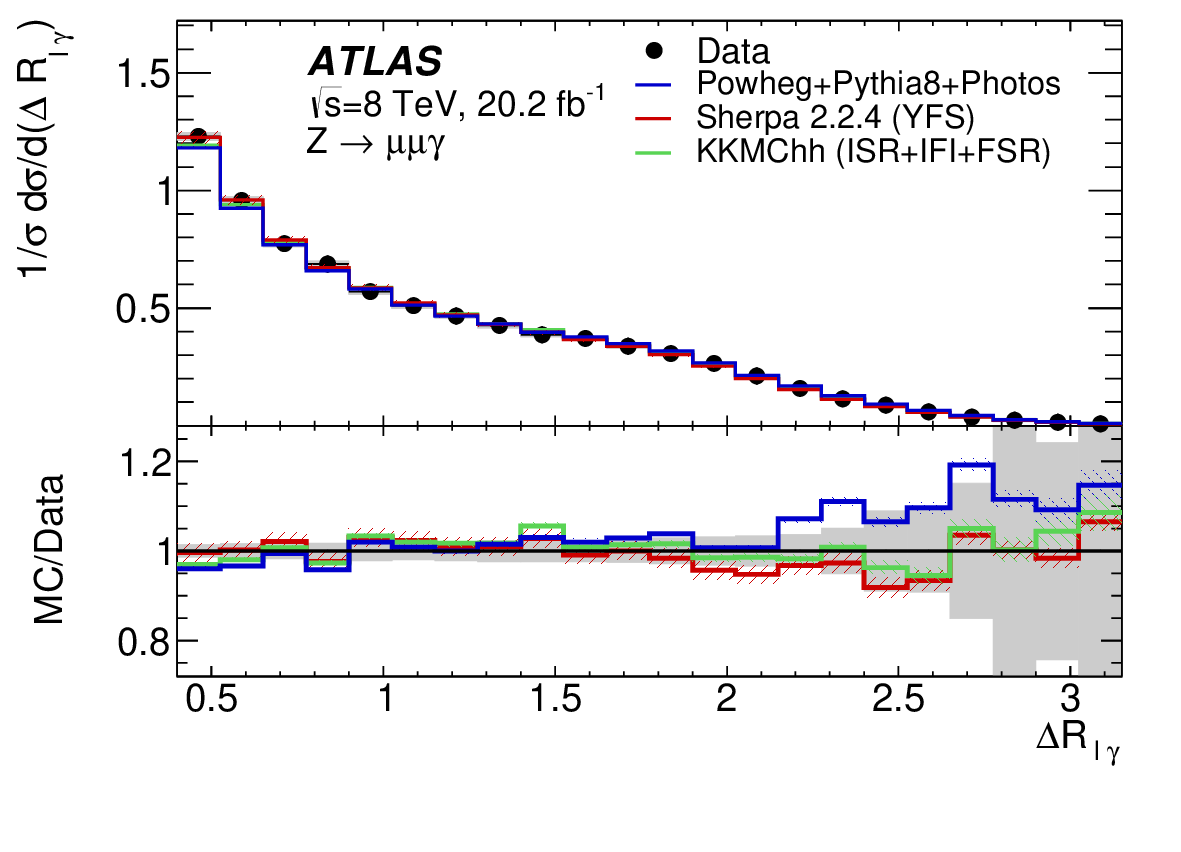}
\includegraphics[width=0.49\columnwidth]{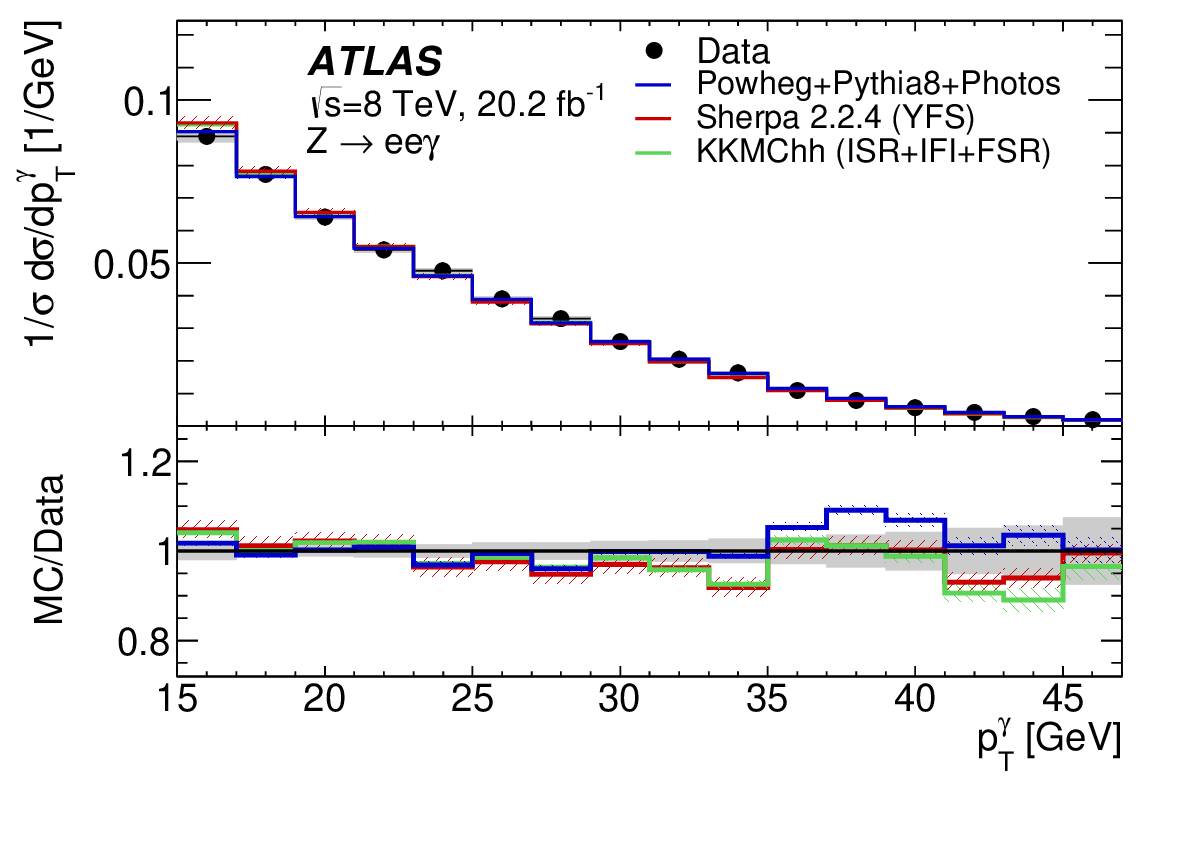}
\hfill
\includegraphics[width=0.49\columnwidth]{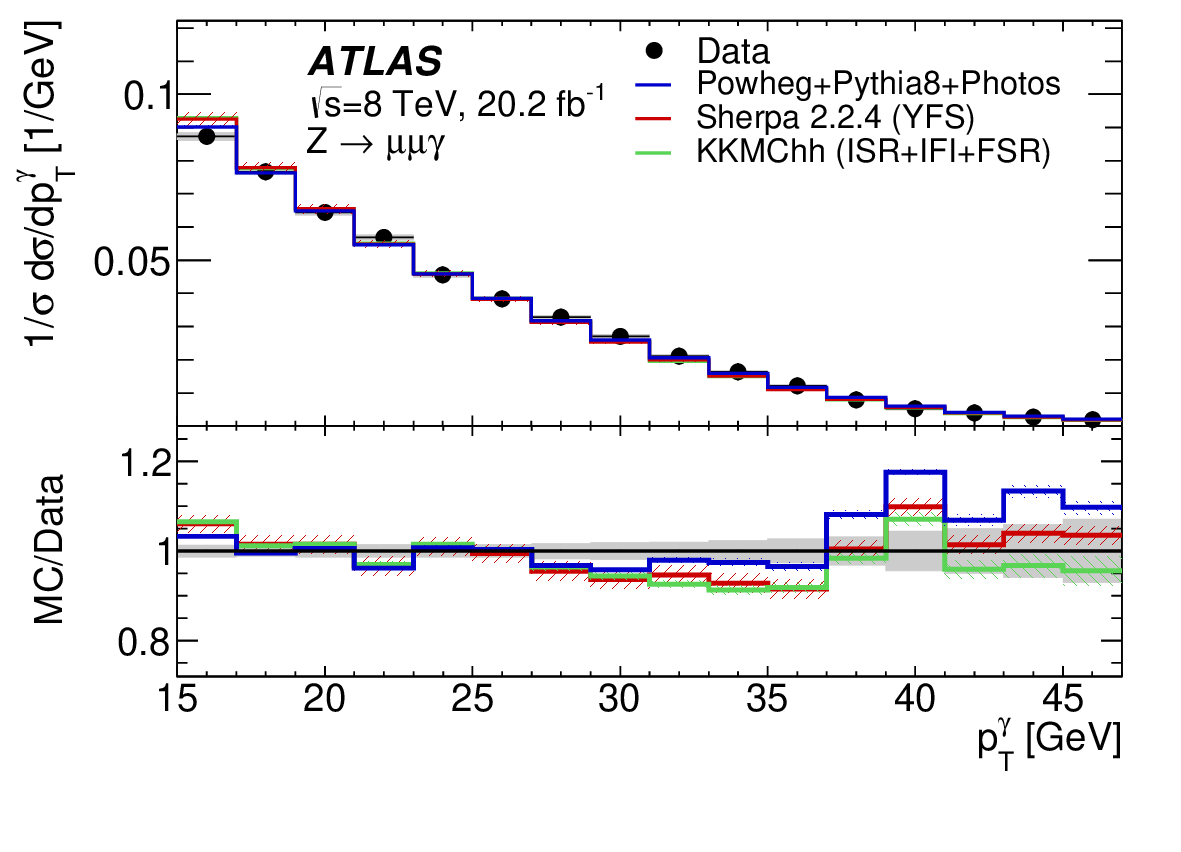}
 
\parbox[t]{1.0\textwidth}{\caption{
Unfolded normalised differential fiducial cross-sections for the \Zeeg~(left) and \Zmmg~(right) channels. The results are shown in the case of bare leptons for the~\mlg~(top), \DeltaRlg~(middle), and \ptg~(bottom) distributions. The data are compared to the predictions from Powheg+Pythia8+Photos, Sherpa~2.2.4, and~KKMChh. The bottom panels represent the ratios of these predictions to the data, where the gray band around unity represents the total data uncertainties and the dashed lines around the predictions represent their respective total uncertainties (see text).
\label{Final20}
}}
\end{center}
\end{figure}

Figure~\ref{Final20} shows in the case of bare leptons, separately for the electron and muon channels, the unfolded normalised differential distributions for the data with their uncertainties compared to the predictions with uncertainties from Powheg+Pythia8+Photos, Sherpa~2.2.4~(YFS), and KKMChh. As discussed in~Section~\ref{sec:samples}, Powheg+Pythia8+Photos considers only QED~FSR and has an associated theoretical uncertainty of~0.2\% obtained by varying the configuration of Photos from its default one to the slightly less precise prediction not including the $Z$-boson matrix-element correction discussed in~Section~\ref{sec:samples}. This small theoretical uncertainty assumes that the contributions from QED~ISR and~IFI are even smaller. The combined QED~ISR/IFI contribution in the fiducial region of the analysis presented in this paper has been estimated with KKMChh to be $3 \times~10^{-3}$ for the integrated fiducial cross-sections, with an uncertainty below~$10^{-3}$. These results have been confirmed using the CompHEP~\cite{CompHEP} code. In the case of the Sherpa~2.2.4 predictions which are also shown in~Fig.~\ref{Final20}, the uncertainty band is much larger and probably over-conservative, about~2\%. This was estimated using a recipe similar to that described in~\cite{SherpaQED}, corresponding to higher-order QED corrections which are only known at the $Z$-boson pole and are not applied outside a very narrow region around~\mZ\ to assess this uncertainty.
 
The agreement between the measurements and the predictions from the MC~simulations is reasonable for all three distributions over most of their range, both for electrons and muons. The agreement between the predictions themselves is better for~\mlg\ than their respective agreement with the data, suggesting perhaps that most of the fluctuations seen in the ratio plots for this observable are related to statistical fluctuations in the data since they are limited to a few bins. However, there is a noticeable discrepancy between the measurement in the electron channel and the Powheg+Pythia8+Photos prediction for values of~\DeltaRlg\ larger than~2.5, at the edge of the phase space where the two leptons from $Z$-boson decay recoil against
the photon.
 
\clearpage

\begin{table}[htb]
\begin{center}
\caption{ \label{SysIntX} Uncertainties in the integrated fiducial cross-section measurements. The values for bare leptons with $m_{ll}>20$ GeV are shown.}
\vspace{5mm}
\begin{tabular}{|l|c|c|}
\hline
Uncertainty & \Zeeg\ & \Zmmg\ \\
\hline
Statistical & 0.7\% &  0.7\% \\
Experimental systematic & 3.5\%  & 2.3\% \\
Luminosity & 1.9\%  & 1.9\%  \\
\hline
QCD theory & 0.3\%& 0.3\% \\
\hline
Total & 4.1\% & 3.1\%\\
\hline
\end{tabular}
\end{center}
\end{table}

\begin{table}[htb]
\begin{center}
\caption{ \label{CrossXr2} Integrated fiducial cross-sections (pb) for
three ranges of the dilepton mass~$m_{ll}$. The measurements are
presented for bare leptons with the breakdown of their uncertainties
together with the predictions from Powheg+Pythia8+Photos normalised as
described in the text.
}
\vspace{5mm}
\begin{tabular}{|l|c|c|} 
\hline
Cross-sections in pb & Measurement & Prediction \\
\hline
\Zeeg\ ($20 < m_{ll} < 80$ GeV) & 3.03 $\pm$ 0.02 (stat) $\pm$ 0.11 (syst)
$\pm$ 0.06 (lumi)  & 2.94 $\pm$ 0.10  \\
\Zmmg\ ($20 < m_{ll} < 80$ GeV) & 3.17 $\pm$ 0.02 (stat) $\pm$ 0.07 (syst)
$\pm$ 0.07 (lumi)  & 3.20 $\pm$ 0.10  \\
\Zeeg\ ($45 < m_{ll} < 80$ GeV) & 2.70 $\pm$ 0.02 (stat) $\pm$ 0.10 (syst)
$\pm$ 0.06 (lumi)  & 2.61 $\pm$ 0.10 \\
\Zmmg\ ($45 < m_{ll} < 80$ GeV) & 2.84 $\pm$ 0.02 (stat) $\pm$ 0.06 (syst)
$\pm$ 0.06 (lumi)  & 2.84 $\pm$ 0.10 \\
\Zeeg\ ($20 < m_{ll} < 45$ GeV) & 0.326 $\pm$ 0.005 (stat) $\pm$ 0.011 (syst) $\pm$ 0.006 (lumi)  & 0.335 $\pm$ 0.01 \\
\Zmmg\ ($20 < m_{ll} < 45$ GeV) & 0.321 $\pm$ 0.005 (stat) $\pm$ 0.006 (syst) $\pm$ 0.006 (lumi) & 0.355 $\pm$ 0.01  \\
\hline
\end{tabular}
\end{center}
\end{table}

\subsection{Integrated fiducial cross-sections}
\label{sec:intXsec}
 
The normalised differential fiducial cross-section measurements presented above are only sensitive to the shape of the distributions, but they benefit from cancellations of experimental systematic uncertainties for the sources that do not vary much with lepton~\pt,~such as the lepton efficiencies. This section presents the integrated fiducial cross-section measurements in the electron and muon channels and discusses the results for three different ranges of the dilepton mass. The uncertainties in the integrated fiducial cross-sections are summarised in Table~\ref{SysIntX}. They include the uncertainty of~1.9\% in the integrated luminosity and are somewhat larger for some of the experimental sources than are those listed in more detail in~Table~\ref{SysTotX} for the reasons explained above.

\begin{table}[htb!]
\begin{center}
\caption{\label{cutsTr2X} Fiducial volume selections for \Zllgg\ events.}
\vspace{5mm}
\begin{tabular}{|c|}
\hline
Two photons with \pT~$>$~15 and 10 GeV, respectively \\
$|\eta_{\gamma}|<2.37$ excluding $1.37<|\eta_{\gamma}|<1.52$ \\
Two same-flavour opposite-sign leptons with \pT~$>$~25,~15~GeV\\
$|\eta_{\mu}| < 2.7$ \\
$|\eta_e| < 2.47$ \\
\DeltaRlg~$> 0.4$ for both photons, \DeltaRgg~$> 0.4$ \\
$m_{ll\gamma} < 80$~GeV for both photons \\
\hline
\end{tabular}
\end{center}
\end{table}

The comparisons between the measured integrated cross-sections over three different dilepton mass ranges and the corresponding theory predictions, normalised to the QCD NNLO prediction used in~\cite{STDM-2016-02} for the total inclusive cross-section for $Z$-boson production, are presented in~Table~\ref{CrossXr2}. The agreement between the measurements and the predictions is reasonable within their comparable overall uncertainties. The uncertainties in the theory predictions are also taken from~\cite{STDM-2016-02}. As explained in~Section~\ref{sec:Intro} and shown in~Table~\ref{CrossXr2} for the theory predictions, the cross-sections for the \Zmmg\ process are expected to be larger than those for the \Zeeg~process. This difference is enhanced significantly in this measurement by the higher acceptance, due to the pseudorapidity coverage, of the \Zmmg~channel with respect to the \Zeeg~channel.

\section{First measurement of the \Zllgg~process}
\label{sec:Z2l2g}

\begin{figure}[htb!]
\begin{center}
\includegraphics[width=0.47\textwidth]{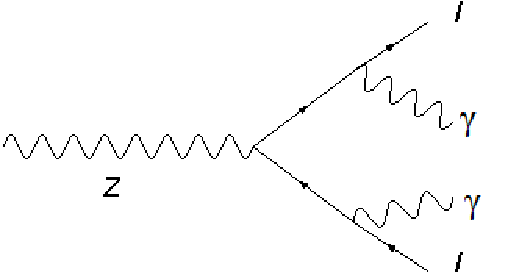}
\hfill
\includegraphics[width=0.47\textwidth]{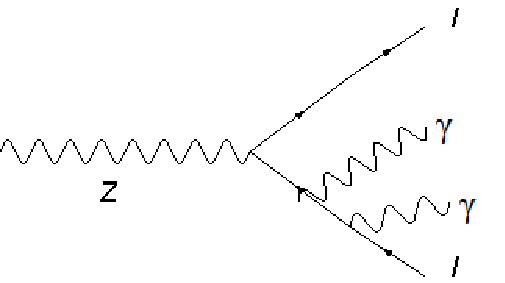}\\
\caption{Examples of $Z \to ll\gamma\gamma$
final-state radiation diagrams.
\label{Fsr2y}}
\end{center}
\end{figure}

\begin{figure}[htb!]
\begin{center}
\includegraphics[width=0.49\columnwidth]{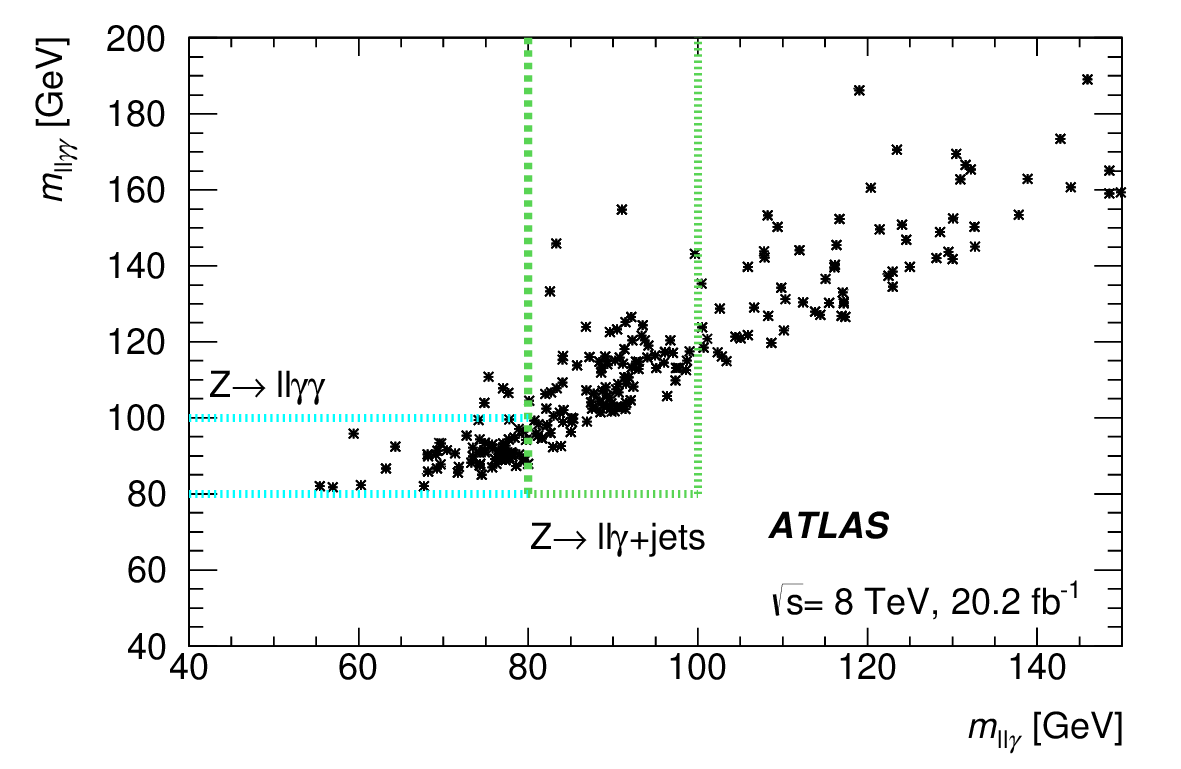}
\hfill
\includegraphics[width=0.49\columnwidth]{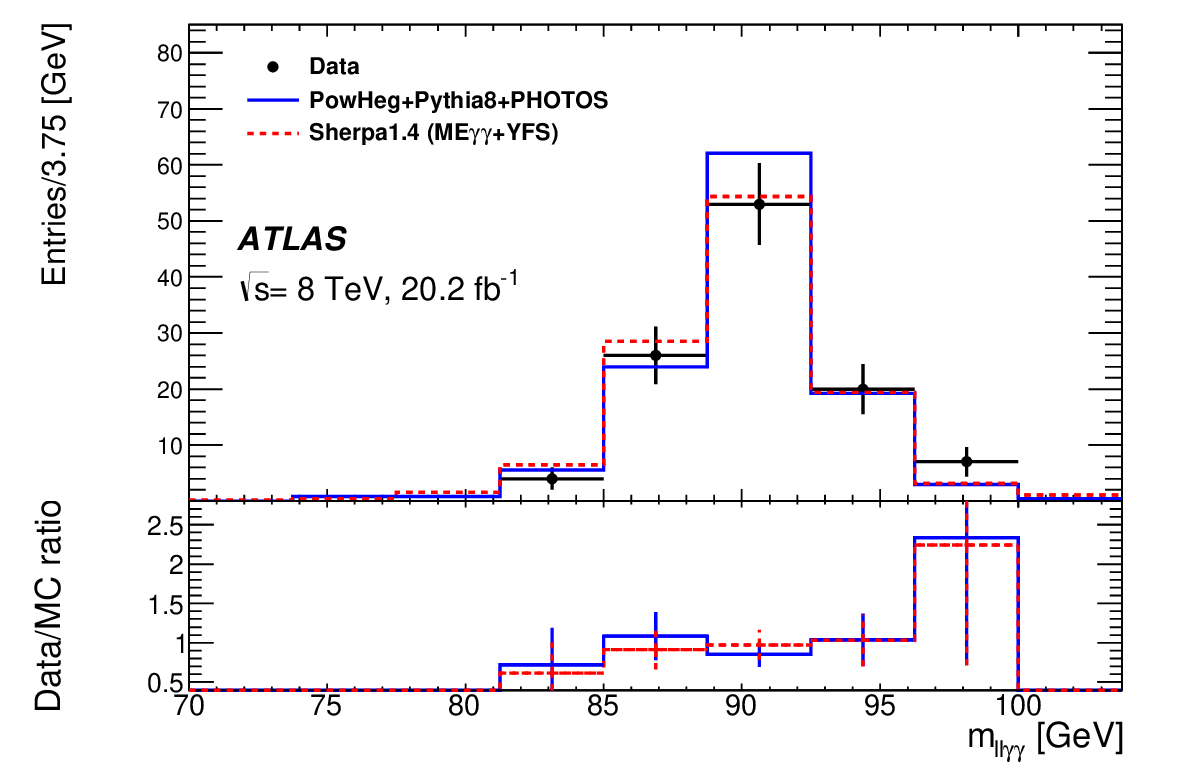}\\
\includegraphics[width=0.49\columnwidth]{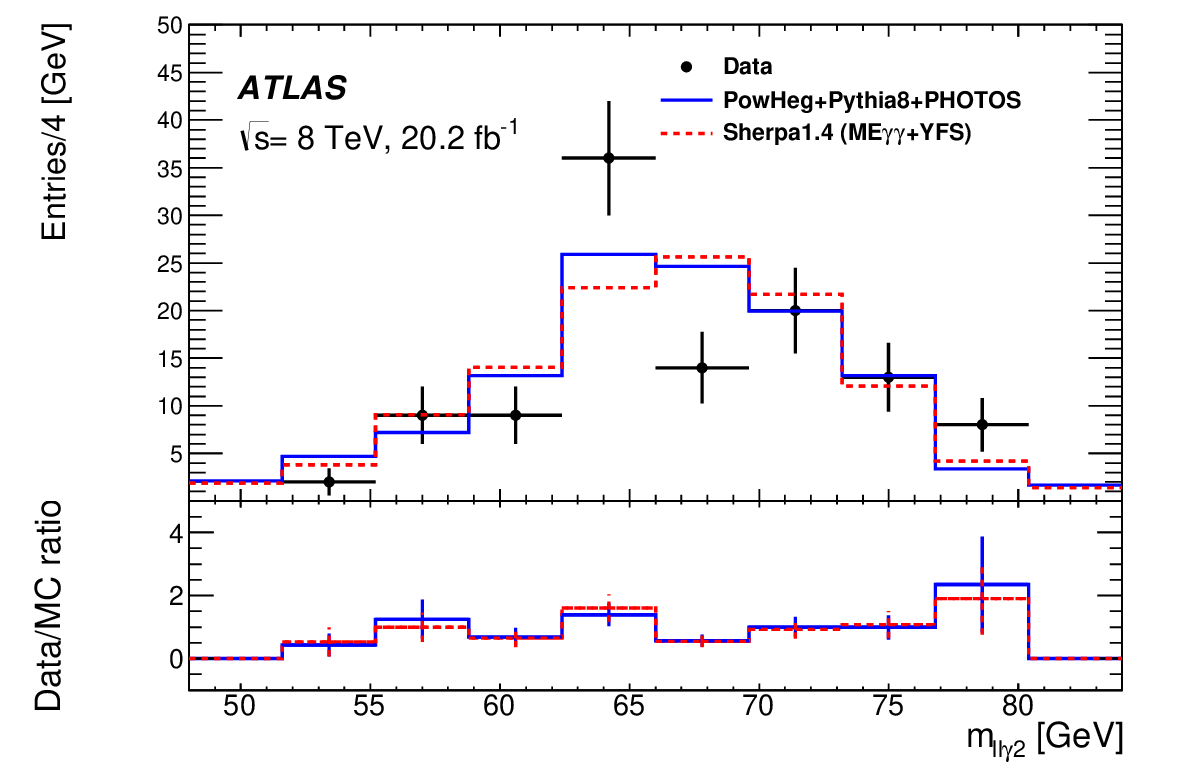}
\hfill
\includegraphics[width=0.49\columnwidth]{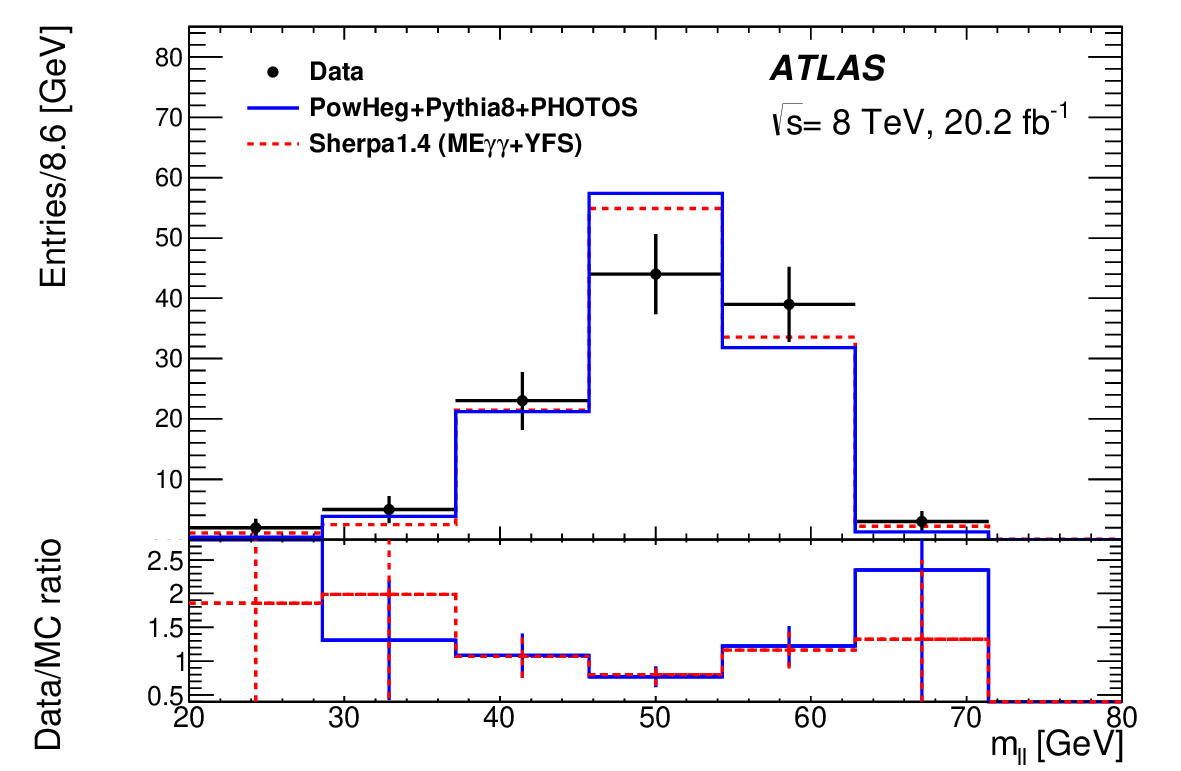}\\
\includegraphics[width=0.49\columnwidth]{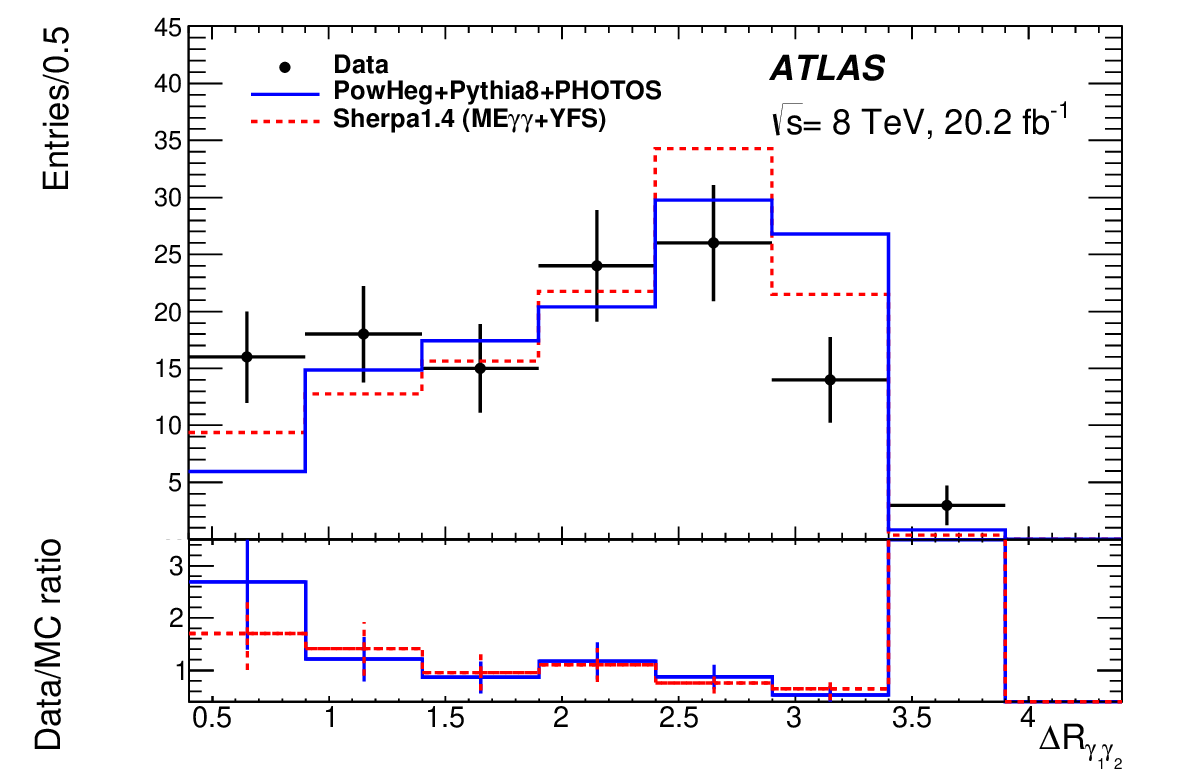}
\hfill
\includegraphics[width=0.49\columnwidth]{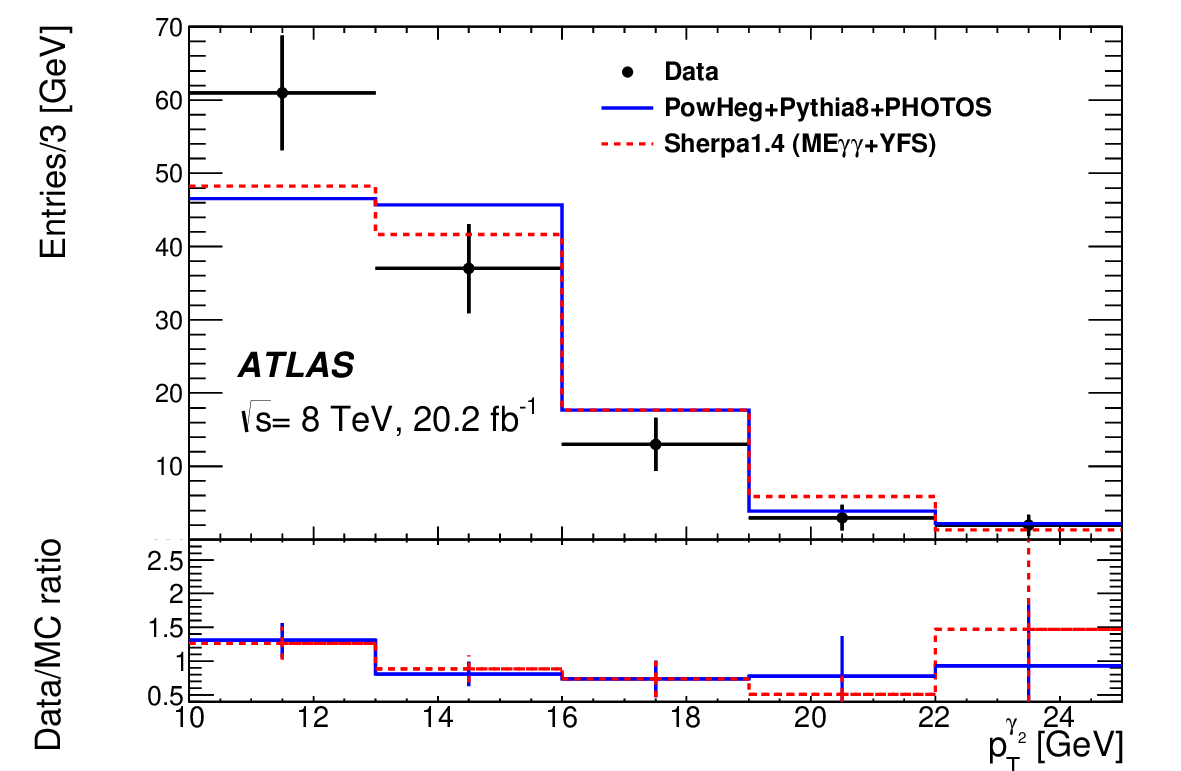}\\
 
\caption{For all selected \Zllgg~candidates before applying any selection on the invariant mass of the two leptons and each photon described in the text, distribution of invariant mass of the two leptons and the two photons,~$m_{ll\gamma\gamma}$ versus the invariant mass of the two leptons and of the highest~\pT photon,~$m_{ll\gamma}$ (top left). The horizontal dashed (blue) rectangle indicates the region where the \Zllgg~signal is expected and the vertical dashed (green) rectangle indicates the region dominated by background from \Zllg~events accompanied by one or more jets.
Also shown after all selection requirements are the distributions of the total invariant mass (top right), of the invariant mass of the two leptons and the second photon, of the dilepton mass, of the angular distance between the two photons, and of the transverse momentum of the second photon. The data are compared to predictions from Powheg+Pythia8+Photos and Sherpa~1.4~ME$_{\gamma \gamma}$, normalised both to the data. The results are shown for bare leptons, and electrons and muons have been combined. The uncertainties shown are statistical.
\label{M2lFor2g0}
}
\end{center}
\end{figure}

This section presents a first measurement of the \Zllgg~process at the
LHC (Fig.~\ref{Fsr2y}). Given the limited statistics expected for the
dataset considered here, the candidate event samples for each charged
lepton flavour were combined together, and the selection requirements
on the sub-leading photon were loosened with respect to the leading
one. The sample considered was the same one as for the \Zllg~analysis
in terms of the charged lepton requirements, with however a tighter
selection requirement of~\pT~$>$~15~GeV for the softer of the two
leptons, but without any requirement on the dilepton mass. The two
photons were required to be within the same pseudorapidity range as
for the \Zllg~analysis with thresholds in~\pT set to respectively
15~and 10~GeV and within an angular distance,~\DeltaRlg\, of at
least~0.4 from each charged lepton. In addition, the angular distance
between the two photons,~\DeltaRgg, was also required to be larger
than~0.4. Figure~\ref{M2lFor2g0}~(top left) shows the invariant mass
of the two leptons and two photons,~$m_{ll\gamma\gamma}$, versus the
invariant mass of the two leptons and the highest~\pT
photon,~$m_{ll\gamma}$. As illustrated in this figure, for values
of~$m_{ll\gamma}$ larger than 80~GeV, significant contributions of
background from \Zllg~+jet(s)  events and from \Zll~+jets events are
expected.
 
In order to reduce these contributions as well as possible
contributions from \Zllg~events with initial state QED radiation to
the percent level or less, the invariant mass,~$m_{ll\gamma}$, was
required to be below~80~GeV for both photons. These requirements are
summarised as fiducial volume selections in~Table~\ref{cutsTr2X}. A
total of 116~events, corresponding to 61 electron pairs and 55 muon
pairs, were thus selected as \Zllgg\ candidates. The dominant
background from~\Zllg+jet events was estimated by applying the same
two-dimensional sideband method as that described for the main
analysis: the quality and isolation criteria were inverted for the
second photon. The background was found to be less than one event and
Fig.~\ref{M2lFor2g0}~(top right) illustrates this through the observed
distribution of~$m_{ll\gamma\gamma}$ which displays no events outside
the pole region.
 
The data were compared to MC~\Zllgg~samples produced using the same
Powheg+Pythia8+Photos version used for the main analysis,
Sherpa~1.4~ME$_{\gamma \gamma}$ (see Section~\ref{sec:samples}),
and~KKMChh. Figure~\ref{M2lFor2g0}~(top right, middle and bottom)
shows that the predictions describe the data within their large
statistical uncertainties for a selection of observables.

Finally, the data were corrected for detector efficiencies and
resolution, and fiducial cross-sections were extracted together with
unfolded distributions of several observables. The fiducial
cross-section is measured to be $22.2 \pm
2.1$~(stat)~$\pm$~1.2~(exp)~$\pm$~4.1~(theo)~fb for bare leptons. In
contrast to the \Zllg~case, the expected efficiencies from the
Sherpa~1.4~ME$_{\gamma \gamma}$ and Powheg+Pythia8+Photos MC~samples
are significantly different and this difference was included as an
additional theoretical systematic uncertainty in the measurement
result for the \Zllgg~cross--section quoted here.
 
\begin{figure}[htb!]
\begin{center}
\includegraphics[width=0.49\columnwidth]{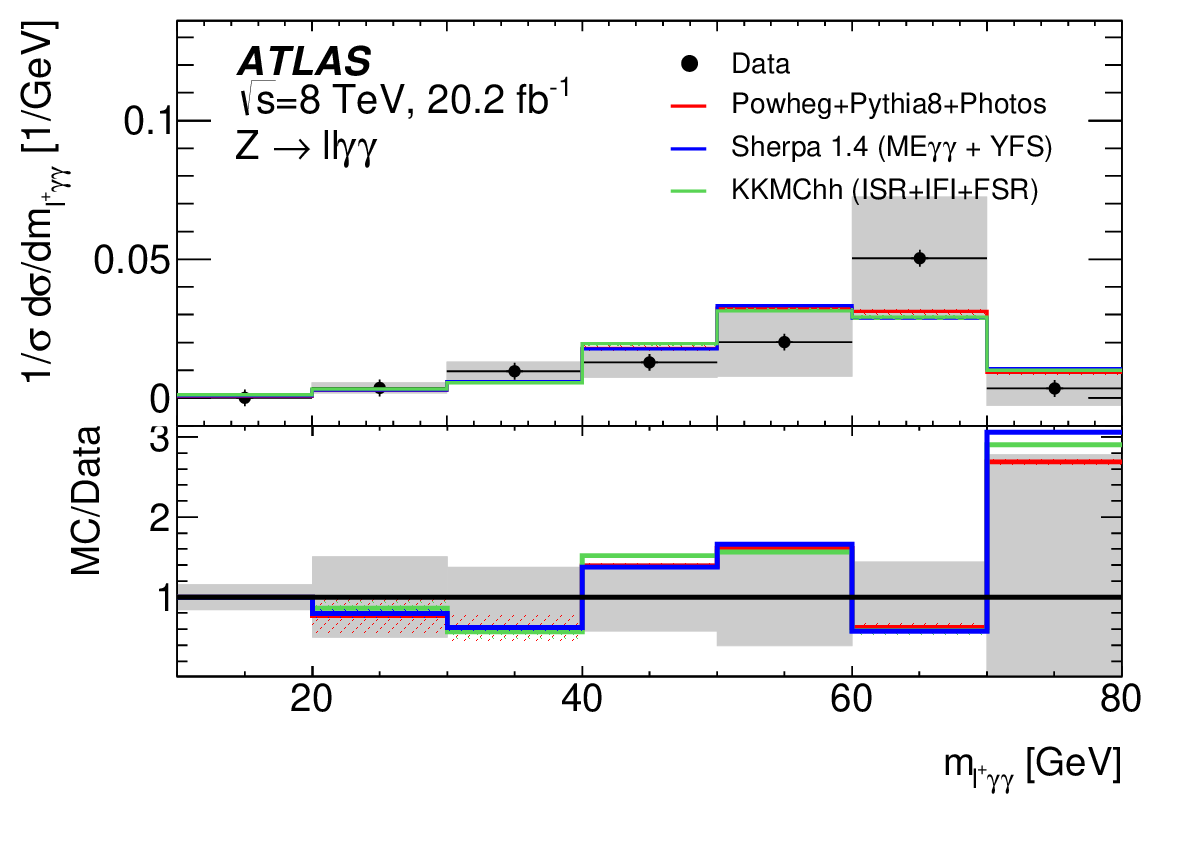}
\hfill
\includegraphics[width=0.49\columnwidth]{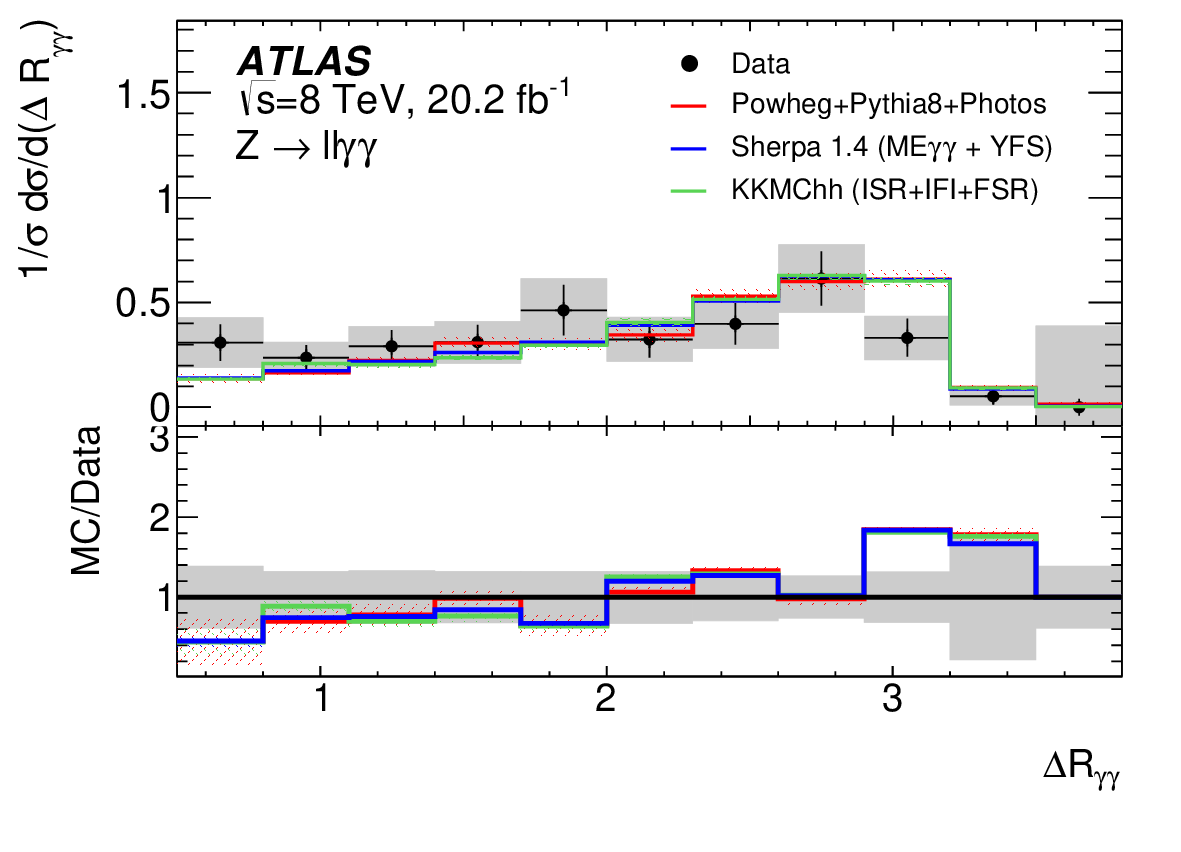}\\
\includegraphics[width=0.49\columnwidth]{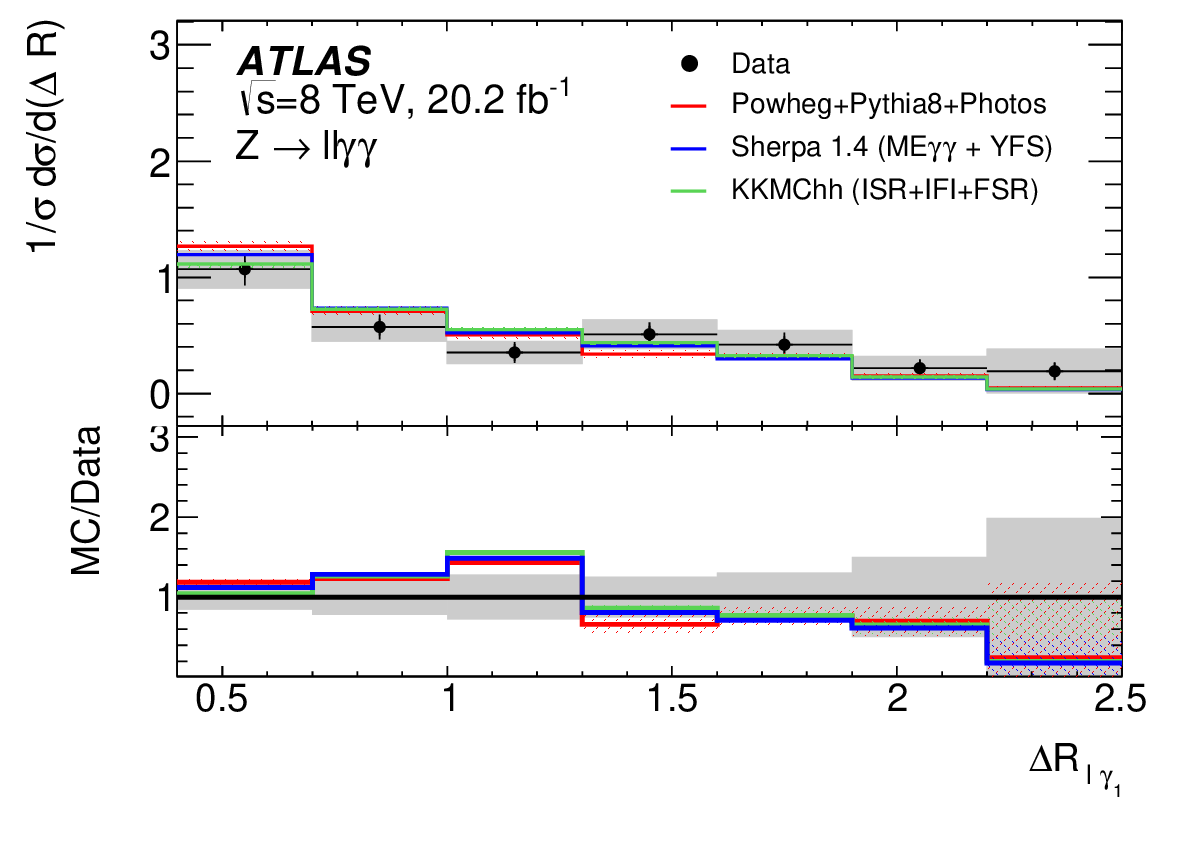}
\hfill
\includegraphics[width=0.49\columnwidth]{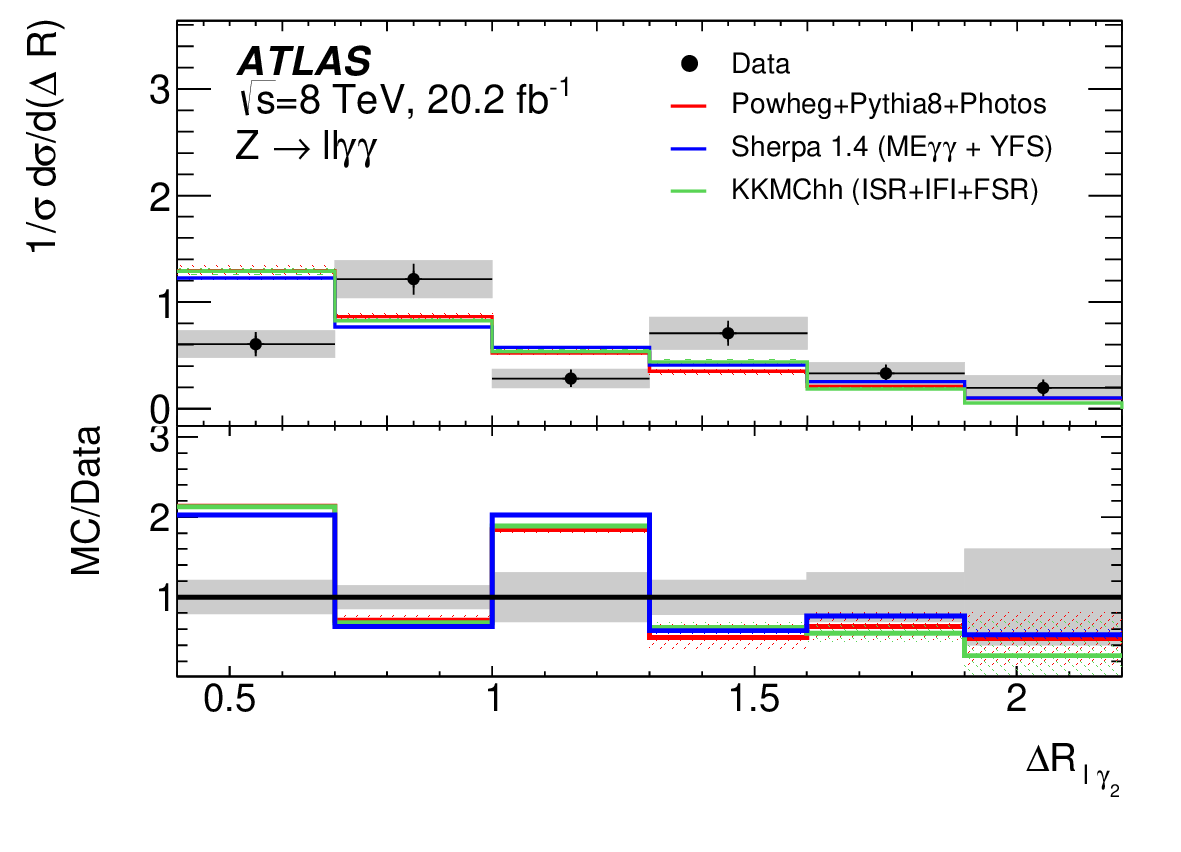}\\
\includegraphics[width=0.49\columnwidth]{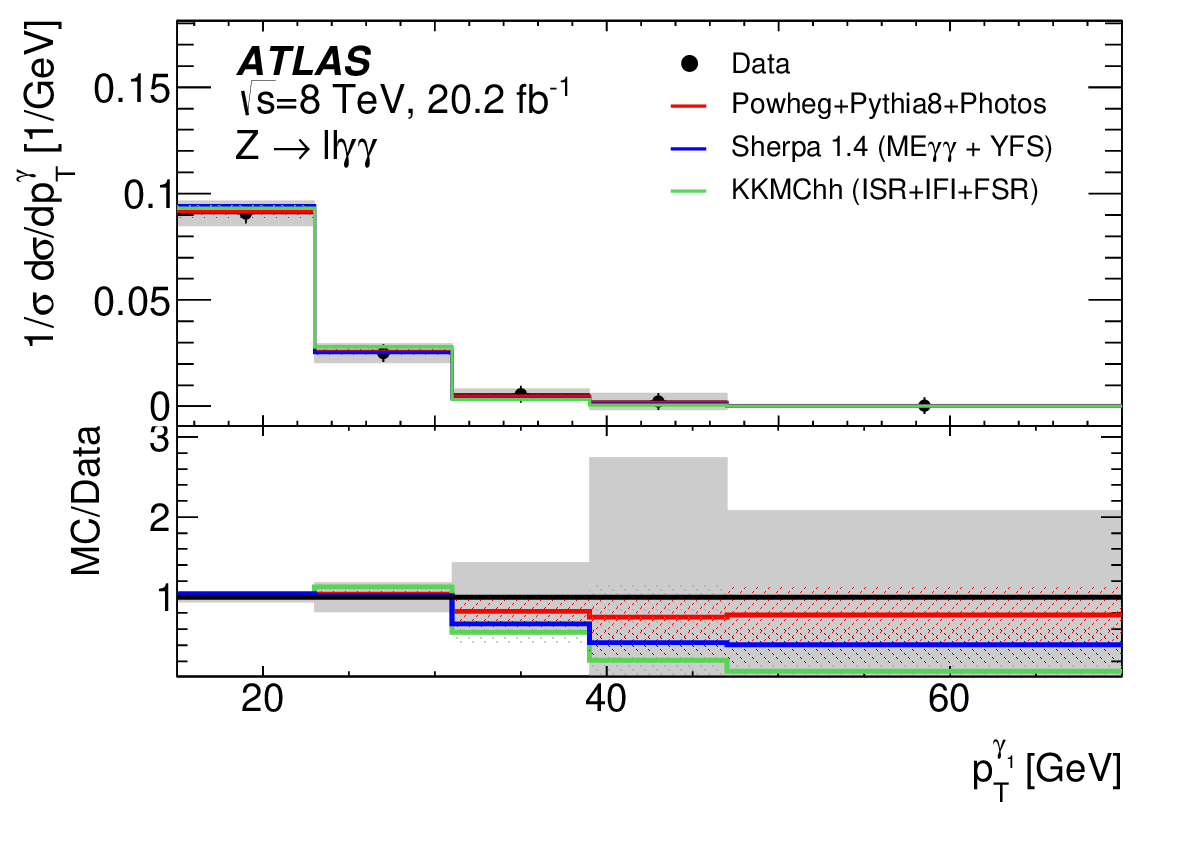}
\hfill
\includegraphics[width=0.49\columnwidth]{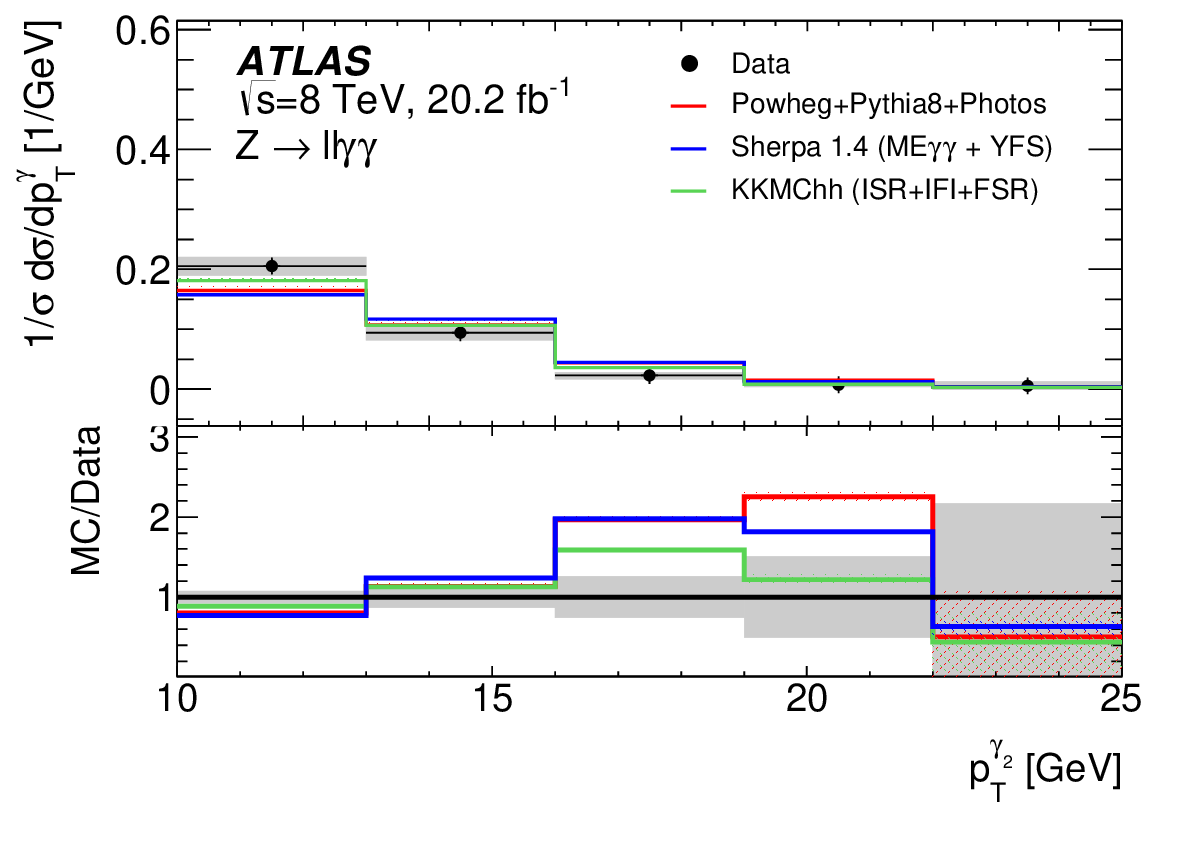}\\
 
\caption{For all selected \Zllgg~candidates, unfolded normalised distributions of $m_{l^+\gamma\gamma}$ (top left), of~\DeltaRgg\ (top right), of the angular distance between the first (middle left) and second (middle right) photon and the closest lepton, and of the transverse momentum spectrum of the two photons (bottom). The unfolded data are compared to predictions from Powheg+Pythia8+Photos, Sherpa~1.4~ME$_{\gamma \gamma}$, and~KKMChh. The results are shown for bare leptons, and electrons and muons have been combined.
The gray band represents the total data uncertainties and
the dashed lines around the predictions represent their statistical uncertainties, which are dominant, for the predictions.
\label{unfoldMlg2gX}}
\end{center}
\end{figure}

The expected
fiducial cross-section from Sherpa~1.4~ME$_{\gamma \gamma}$ is $9.9
\pm 0.1$~(stat)~fb, that from Powheg+Pythia8+Photos is $9.4 \pm
0.3$~(stat)~fb. Both values are significantly smaller than the
measured one. However, the expected fiducial cross-section from
Sherpa~2.2.4 is $20.9 \pm 0.2$~(stat)~fb, in agreement with the data.
Although the purely QED systematic uncertainty in the predictions
cannot explain the observed discrepancies between the measurement and
two of the predictions, the overall systematic uncertainty in the
predictions, which should include QCD and mixed QCD/QED effects, is
not known. This uncertainty is clearly of QCD~origin, as shown by the
two versions of~Sherpa. Figure~\ref{unfoldMlg2gX} shows the unfolded
distributions for the data compared to the predictions.
The uncertainties shown are the combined statistical and systematic
uncertainties for the data and only the statistical uncertainties for
the predictions. The agreement in shape between data and predictions
is reasonable within the uncertainties of this first measurement.
 
\clearpage

\section{Conclusions}
 
This paper presents the first ATLAS measurement of unfolded normalised differential fiducial cross-sections for \Zeeg\ and \Zmmg\ final states together with a few integrated fiducial cross-sections. The overall accuracy of the differential measurements is at the 1-2\%~level over a large fraction of the overall phase space, with an average overall systematic uncertainty of~0.6\% for the chosen observables, \mlg, \DeltaRlg, and~\ptg.
 
The results are in agreement with predictions of MC~generators containing state-of-the-art QED~FSR calculations such as Photos (in PowHeg+Pythia8+Photos) and YFS (in Sherpa 2.2.4). They are also in agreement with the recent calculations of KKMChh which provides a consistent and gauge-invariant breakdown of the different contributions to final states containing two charged leptons and a high-\pt\ photon, and thus demonstrates that the QED~ISR and~IFI contributions are very small for the fiducial selections of the analysis described in this paper.
 
Finally, the \Zllgg\ process has been observed for the first time.
The differential distributions for this process are described by the
available calculations within present large statistical uncertainties
of the measurement. The \Zllgg\ fiducial cross-section is consistent with Sherpa
2.2.4 and is twice as large as predicted by PowHeg+Pythia8+Photos and
Sherpa 1.4~ME$_{\gamma \gamma}$.

\clearpage
 
\section*{Acknowledgements}


We thank CERN for the very successful operation of the LHC, as well as the
support staff from our institutions without whom ATLAS could not be
operated efficiently.
 
We acknowledge the support of
ANPCyT, Argentina;
YerPhI, Armenia;
ARC, Australia;
BMWFW and FWF, Austria;
ANAS, Azerbaijan;
CNPq and FAPESP, Brazil;
NSERC, NRC and CFI, Canada;
CERN;
ANID, Chile;
CAS, MOST and NSFC, China;
Minciencias, Colombia;
MEYS CR, Czech Republic;
DNRF and DNSRC, Denmark;
IN2P3-CNRS and CEA-DRF/IRFU, France;
SRNSFG, Georgia;
BMBF, HGF and MPG, Germany;
GSRI, Greece;
RGC and Hong Kong SAR, China;
ISF and Benoziyo Center, Israel;
INFN, Italy;
MEXT and JSPS, Japan;
CNRST, Morocco;
NWO, Netherlands;
RCN, Norway;
MEiN, Poland;
FCT, Portugal;
MNE/IFA, Romania;
MESTD, Serbia;
MSSR, Slovakia;
ARRS and MIZ\v{S}, Slovenia;
DSI/NRF, South Africa;
MICINN, Spain;
SRC and Wallenberg Foundation, Sweden;
SERI, SNSF and Cantons of Bern and Geneva, Switzerland;
MOST, Taiwan;
TENMAK, T\"urkiye;
STFC, United Kingdom;
DOE and NSF, United States of America.
In addition, individual groups and members have received support from
BCKDF, CANARIE, Compute Canada and CRC, Canada;
PRIMUS 21/SCI/017 and UNCE SCI/013, Czech Republic;
COST, ERC, ERDF, Horizon 2020 and Marie Sk{\l}odowska-Curie Actions, European Union;
Investissements d'Avenir Labex, Investissements d'Avenir Idex and ANR, France;
DFG and AvH Foundation, Germany;
Herakleitos, Thales and Aristeia programmes co-financed by EU-ESF and the Greek NSRF, Greece;
BSF-NSF and MINERVA, Israel;
Norwegian Financial Mechanism 2014-2021, Norway;
NCN and NAWA, Poland;
La Caixa Banking Foundation, CERCA Programme Generalitat de Catalunya and PROMETEO and GenT Programmes Generalitat Valenciana, Spain;
G\"{o}ran Gustafssons Stiftelse, Sweden;
The Royal Society and Leverhulme Trust, United Kingdom.
 
The crucial computing support from all WLCG partners is acknowledged gratefully, in particular from CERN, the ATLAS Tier-1 facilities at TRIUMF (Canada), NDGF (Denmark, Norway, Sweden), CC-IN2P3 (France), KIT/GridKA (Germany), INFN-CNAF (Italy), NL-T1 (Netherlands), PIC (Spain), ASGC (Taiwan), RAL (UK) and BNL (USA), the Tier-2 facilities worldwide and large non-WLCG resource providers. Major contributors of computing resources are listed in Ref.~\cite{ATL-SOFT-PUB-2023-001}.


\clearpage
\printbibliography
 
\clearpage

\clearpage
\input{atlas_authlist}

\end{document}

%% file: atlas_authlist.tex
 
\begin{flushleft}
\hypersetup{urlcolor=black}
{\Large The ATLAS Collaboration}

\bigskip

\AtlasOrcid[0000-0002-6665-4934]{G.~Aad}$^\textrm{\scriptsize 101}$,
\AtlasOrcid[0000-0002-5888-2734]{B.~Abbott}$^\textrm{\scriptsize 119}$,
\AtlasOrcid[0000-0002-7248-3203]{D.C.~Abbott}$^\textrm{\scriptsize 102}$,
\AtlasOrcid[0000-0002-1002-1652]{K.~Abeling}$^\textrm{\scriptsize 55}$,
\AtlasOrcid[0000-0002-8496-9294]{S.H.~Abidi}$^\textrm{\scriptsize 29}$,
\AtlasOrcid[0000-0002-9987-2292]{A.~Aboulhorma}$^\textrm{\scriptsize 35e}$,
\AtlasOrcid[0000-0001-5329-6640]{H.~Abramowicz}$^\textrm{\scriptsize 150}$,
\AtlasOrcid[0000-0002-1599-2896]{H.~Abreu}$^\textrm{\scriptsize 149}$,
\AtlasOrcid[0000-0003-0403-3697]{Y.~Abulaiti}$^\textrm{\scriptsize 116}$,
\AtlasOrcid[0000-0003-0762-7204]{A.C.~Abusleme~Hoffman}$^\textrm{\scriptsize 136a}$
% ...
\bigskip
\\

% ...
$^{29}$Physics Department, Brookhaven National Laboratory, Upton NY; United States of America.\\
% ...
$^{35}$$^{(a)}$Facult\'e des Sciences Ain Chock, R\'eseau Universitaire de Physique des Hautes Energies - Universit\'e Hassan II, Casablanca;$^{(b)}$Facult\'{e} des Sciences, Universit\'{e} Ibn-Tofail, K\'{e}nitra;$^{(c)}$Facult\'e des Sciences Semlalia, Universit\'e Cadi Ayyad, LPHEA-Marrakech;$^{(d)}$LPMR, Facult\'e des Sciences, Universit\'e Mohamed Premier, Oujda;$^{(e)}$Facult\'e des sciences, Universit\'e Mohammed V, Rabat;$^{(f)}$Institute of Applied Physics, Mohammed VI Polytechnic University, Ben Guerir; Morocco.\\
% ...
$^{55}$II. Physikalisches Institut, Georg-August-Universit\"{a}t G\"ottingen, G\"ottingen; Germany.\\
% ...
$^{101}$CPPM, Aix-Marseille Universit\'e, CNRS/IN2P3, Marseille; France.\\
$^{102}$Department of Physics, University of Massachusetts, Amherst MA; United States of America.\\
% ...
$^{116}$Department of Physics, New York University, New York NY; United States of America.\\
% ...
$^{119}$Homer L. Dodge Department of Physics and Astronomy, University of Oklahoma, Norman OK; United States of America.\\
% ...
$^{136}$$^{(a)}$Departamento de F\'isica, Pontificia Universidad Cat\'olica de Chile, Santiago;$^{(b)}$Millennium Institute for Subatomic physics at high energy frontier (SAPHIR), Santiago;$^{(c)}$Instituto de Investigaci\'on Multidisciplinario en Ciencia y Tecnolog\'ia, y Departamento de F\'isica, Universidad de La Serena;$^{(d)}$Universidad Andres Bello, Department of Physics, Santiago;$^{(e)}$Instituto de Alta Investigaci\'on, Universidad de Tarapac\'a, Arica;$^{(f)}$Departamento de F\'isica, Universidad T\'ecnica Federico Santa Mar\'ia, Valpara\'iso; Chile.\\
% ...
$^{149}$Department of Physics, Technion, Israel Institute of Technology, Haifa; Israel.\\
$^{150}$Raymond and Beverly Sackler School of Physics and Astronomy, Tel Aviv University, Tel Aviv; Israel.\\
% ...
$^{a}$ Also Affiliated with an institute covered by a cooperation agreement with CERN.\\
$^{b}$ Also at An-Najah National University, Nablus; Palestine.\\
$^{c}$ Also at Borough of Manhattan Community College, City University of New York, New York NY; United States of America.\\
$^{d}$ Also at Bruno Kessler Foundation, Trento; Italy.\\
$^{e}$ Also at Center for High Energy Physics, Peking University; China.\\
$^{f}$ Also at Center for Interdisciplinary Research and Innovation (CIRI-AUTH), Thessaloniki; Greece.\\
% ...

\end{flushleft}
